\documentclass[useAMS]{mn2e}

\usepackage{amsmath}
\newcommand{\dfn}[3]{\dfrac{\mathrm{d}^{#3} #1}{\mathrm{d} #2^{#3}}}
\newcommand{\pdn}[3]{\dfrac{\partial^{#3} #1}{\partial #2^{#3}}}
\newcommand{\rd}{\mathrm{d}}

\usepackage{subfigure}
\usepackage[dvips]{graphicx}
\usepackage{threeparttable}

\def\pmb#1{\mbox{\boldmath$#1$}}
\def\pmbmt#1{\pmb{\sf #1}}

\def\be{\begin{eqnarray}}
\def\ee{\end{eqnarray}}

\begin{document}
\title[Stars having purely toroidal magnetic field]
  {Non-radial oscillations of the magnetized rotating stars with purely toroidal magnetic fields}
\author[H. Asai, U. Lee, and S. Yoshida]
  {Hidetaka Asai, Umin Lee, and Shijun Yoshida \\
Astronomical Institute, Tohoku University, Sendai 980-8578, Japan}

\date{Received / Accepted}
\pubyear{0000} \volume{000} \pagerange{000-000} \onecolumn

\maketitle \label{firstpage}
 \nokeywords

\begin{abstract}
We calculate non-axisymmetric oscillations of uniformly rotating
polytropes magnetized with a purely toroidal magnetic field, taking account of
the effects of the deformation due to the magnetic field. 
As for rotation, we consider only the effects of Coriolis force on the oscillation modes, ignoring 
those of the centrifugal force, that is, of the rotational deformation of the star.
Since separation of variables is not possible for the
oscillation of rotating magnetized stars, we employ finite series
expansions for the perturbations using spherical harmonic
functions. 
We calculate magnetically modified normal modes such as $g$-,
$f$-, $p$-, $r$-, and inertial modes. 
In the lowest order, the frequency shifts produced by the magnetic field scale
with the square of the characteristic Alfv\'en frequency.
As a measure of the effects of the magnetic field, we calculate the proportionality constant for the frequency shifts for
various oscillation modes.
We find that the effects of the deformation are significant for
high frequency modes such as $f$- and $p$-modes but unimportant for low frequency modes
such as $g$-, $r$-, and inertial modes. 
\end{abstract}


\section{Introduction}

Since the first discovery report of quasi-periodic oscillations (QPOs) 
in the tail of the giant X/$\gamma$-ray flare from the soft $\gamma$-ray repeater (SGR) 1806-20 (Israel et al. 2005), 
extensive theoretical studies have been carried out to identify physical mechanisms responsible for
the QPOs.
SGRs belong to what we call magnetars, neutron stars possessing an extremely strong magnetic field
as strong as $10^{15}$G at the surface.
Giant flares have so far been observed only from three SGRs, that is, SGR 0526-66, 1900+44, and
1806-20, and just once for each of the SGRs, indicating
that giant flares from magnetars are quite rare events.
For magnetars, see reviews, for example, by Woods \& Thompson (2006) and Mereghetti (2008).
Analyzing archival data of another magnetar candidate SGR 1900+14, 
Strohmayer \& Watts (2005) have succeeded in identifying QPOs in the X-ray giant flare that was observed in 1998.
QPOs frequencies now identified in the giant flares from the two magnetar candidates are 
18, 30, 92.5, 150, 626 Hz for SGR 1806-20 
(Israel et al. 2005; Strohmayer \& Watts 2006; Watts \& Strohmayer 2006), and 
28, 53.5, 84, 155 Hz for SGR 1900+14 (Strohmayer \& Watts 2005).
Employing Bayesian statistics, Hambaryan et al. (2011) have reanalyzed the data for SGR 1806-20 to identify
QPO frequencies at 16.9, 21.4, 36.8, 59.0, 61.3, and 116.3 Hz.
For the giant flare in 1979
from SGR 0526-66, Watts (2011) mentioned in her review paper a report of a QPO at $\sim 43$Hz, 
but she also suggested the difficulty in the frequency analysis in the impulsive phase of
the burst.
Here, it is worth mentioning a promising recent attempt to find QPOs in short recurrent bursts in
SGRs.
Huppenkothen et al. (2014) have succeeded in identifying candidate signals at
260, 93, and 127 Hz from J1550-5418, where they used Bayesian statistics for the analysis.

The QPOs are now regarded
as a manifestation of global oscillations of the underlying neutron
stars, and it is expected that they can be used for
seismological studies of the magnetars.
Seismological studies of magnetars may have started with a paper by Duncan (1998),
who suggested that frequent starquakes in SGRs could excite crustal toroidal modes and 
the burst emission modulated at the mode frequencies would be detected.
The detection of QPOs in the giant flare from SGR 1806-20 in 2004 (Israel et al. 2005) has given a huge trigger
leading to subsequent intensive theoretical studies of QPOs in magnetars.
In the early studies of QPOs in magnetars (e.g., Piro 2005; Lee 2007), 
the oscillations were assumed to be practically confined in the solid crust, as Duncan (1998) anticipated, and 
the effects of a magnetic field in the fluid core were ignored.
Since magnetars are believed to possess an extremely strong magnetic
field threading both the solid crust and the fluid core, to determine the oscillation frequency spectra
of the stars, we need to correctly take account of the effects of the
strong magnetic field in both regions on the oscillations. 
Applying a toy model,
Glampedakis, Samuelsson, and Andersson (2006) discussed toroidal oscillations
as global discrete modes residing in the fluid core and in the solid crust both threaded by a strong magnetic field,
and showed that the modes that are most likely to be excited in magnetars are such that the crust and
the core oscillate in concert.
Also using a toy model, Levin (2006, 2007) put forward a different idea that Alfv\'en modes in the fluid core may lead 
to the formation of continuum frequency spectra and that toroidal crust modes will be rapidly damped as a result of 
resonant absorption in the core, and even suggested that there exist no discrete normal modes for the strongly magnetized
neutron stars.
However, assuming a pure poloidal field threading both the solid crust and the fluid core, 
Lee (2008) and Asai \& Lee (2014) carried out normal mode calculations of axisymmetric toroidal modes and found
discrete toroidal modes.
Later on, van Hoven \& Levin (2011, 2012), employing spectral method of calculation, have succeeded in 
suggesting the existence of discrete modes in the gaps between continuum frequency bands.
In general relativistic frame work, Sotani et al. (2007) computed normal modes of magnetized neutron stars in the weak magnetic field limit.

Oscillations in magnetized stars are governed by a set of linearized partial differential equations.
Normal mode analysis of the oscillations of 
magnetized stars are not necessarily easy to conduct, partly because 
separation of variables between the radial and the angular
coordinates is in general impossible to represent the perturbations,
and partly because the possible existence of continuum bands in the frequency spectra make it difficult to properly calculate oscillation modes, particularly belonging to the continua (see, e.g., Goedbloed \& Poedts 2004).
In normal mode calculations, we usually employ series expansions of a finite length to represent the perturbations 
so that the 
set of linearized partial differential equations reduces to a set of linear ordinary differential equations for the expansion coefficients.
This process can be very cumbersome when we try to carry out modal analyses for various configurations of 
magnetic fields.
This may be one of the reasons why
MHD simulations have been used to investigate the small amplitude oscillations of magnetized neutron 
stars by many authors including, e.g., Sotani, Kokkotas \& Stergioulas (2008),
Cerd\'a-Dur\'an et al. (2009), Colaiuda \&
Kokkotas (2011, 2012), Gabler et al. (2011, 2012, 2013a,b), Lander et al. (2010), Passamonti \& Lander (2013, 2014).
In the analyses with MHD simulations, 
QPOs are believed to be associated with continuum spectra and 
should be properly distinguished from discrete normal modes
by closely watching motions and phases of various points in the interior.

Configurations of magnetic fields in magnetars are highly uncertain (e.g., Thompson \& Duncan 1993, 1996;
Thompson, Lyuitikov \& Kulkarni 2002).
As shown by the core-collapse supernova MHD simulations (e.g., Kotake, Sato \& Takahashi 2006), 
toroidal fields can be easily produced and amplified when 
winding of the initial seed poloidal fields is effective in the differentially rotating core
even if there is no initial toroidal fields.
For modal analyses, it is desirable to examine various magnetic field configurations.
Most of the modal analyses so far carried out have been for a purely poloidal magnetic field
(Lee 2007, 2008; Sotani et al 2007, 2008; Cerd\'a-Dur\'an et al. 2009; Colaiuda \&
Kokkotas 2011, 2012; Gabler et al. 2011, 2012; Passamonti \& Lander 2013, 2014; Asai \& Lee 2014).
Recently, however, some authors, using MHD simulations, started investigating small amplitude oscillations 
for a purely toroidal magnetic field (Lander et al. 2010; Passamonti \& Lander 2013),
and even for mixed poloidal and toroidal field configurations (Gabler et al. 2013).
Using MHD simulations, for example,
Lander et al (2010), calculated rotational modes ($r$-modes and inertial modes)
of magnetized stars and showed that $r$-modes at rapid rotation tend to
magnetic modes in the slow rotation limit.

In this paper we carry out normal mode analysis of polytropic models with a purely toroidal magnetic field 
for various non-axisymmetri oscillation modes, including $p$-, $f$-, $g$-modes, and rotational modes such as
$r$-modes and inertial modes, where we include the effects of equilibrium deformation due to the magnetic field on the
oscillations.
To calculate rotational modes, we only take account of Coriolis force and include no effects of the centrifugal force.
The oscillation equations for magnetically deformed rotating stars are 
derived by following the formulation similar to that
by Saio (1981) (see also Lee 1993; Yoshida \& Lee 2000a). The numerical method to compute
normal modes of magnetized rotating stars is the same as that in Lee
(2005) (see also Lee 2007), who employed series expansions of a finite
length in terms of spherical harmonic functions for the
perturbations.
This paper is organized as follows. \S 2 describes the method used to
construct a magnetically deformed equilibrium stellar model, and
perturbation equations for non-axisymmetric oscillation modes in
magnetized rotating stars are derived in \S 3. Numeical results are
summarized in \S 4 and we conclude in \S 5. The details of the
oscillation equations solved in this paper and suitable boundary
conditions imposed at the stellar center and surface are given in Appendix~A.

\section{Equilibrium model}

We consider the oscillations of uniformly rotating polytropes with purely toroidal magnetic fields. 
Equilibrium structures of stars having purely toroidal magnetic fields have so far been studied with non-perturbative approaches within the framework of Newtonian 
mechanics (Miketinac 1973)  and of general relativity (Kiuchi \& Yoshida 2008; Frieben \& Rezzolla 2012). 
In this study, we employ a perturbative approach to
construct stars deformed by a purely toroidal magnetic field. 
Following Miketinac (1973), a purely toroidal 
magnetic field  imposed on the stars in equilibrium is assumed to be given by 
\begin{eqnarray}
B_r=0,\quad B_\theta=0,\quad B_\phi=k\rho r\sin\theta,
\end{eqnarray}
where $k\equiv B_0/(\sqrt{2}\rho_cR)$ is a constant, $B_0$ is the parameter used for the strength of the magnetic field in the interior, 
$\rho_c$ is the density at the stellar center, and $R$ is the radius of the star. 
Here, we use spherical polar coordinates $(r,\theta,\phi)$. 
The magnitude of the magnetic field is given by 
$|\pmb{B}|=|B_\phi|=(B_0/\sqrt{2})\hat{\rho}x\sin\theta$, where $x=r/R$ and $\hat{\rho}=\rho/\rho_c$. The fluid velocity $\pmb{v}$ in equilibrium is assumed to be given by 
\begin{eqnarray}
v_r=0,\quad v_\theta=0,\quad v_\phi=r\sin\theta\,\Omega,
\end{eqnarray}
where $\Omega$ denotes the angular velocity of the uniformly rotating star.  In this study, the deformation of the star is assumed to be solely caused by the magnetic fields and the effects of the centrifugal force are ignored. 
By the assumptions (1) and (2), the induction 
and continuity equations are automatically satisfied and need not be considered any further. 
For the toroidal field (1), we can write the Lorentz force per unit mass 
as a potential force, that is,
\be
{1\over 4\pi\rho}(\nabla\times\pmb{B})\times\pmb{B} =-\nabla\left(\frac{B_0^2}{8\pi\rho_c}\hat{\rho}x^2\sin^2\theta\right).
\ee
The structure of a star in equilibrium is then determined by the hydrostatic equation, the Poisson equation, and the equation of state:
\begin{eqnarray}
\nabla p=-\rho\nabla\Psi,
\label{eq:hydro}
\end{eqnarray}
\begin{eqnarray}
\nabla^2\Phi=4\pi G\rho,
\label{eq:poisson}
\end{eqnarray}
\begin{eqnarray}
p=K_c\rho^{1+1/n},
\label{eq:poly}
\end{eqnarray}
where $n$ and $K_c$ are the polytropic index and the structure constant
given by the mass and the radius of the star, $G$ is the
gravitational constant, $\Phi$ is the gravitational potential, and
$\Psi$ is the effective potential defined by
\begin{eqnarray}
\Psi
=\Phi+\frac{1}{3}\omega_A^2r^2\hat{\rho}\left[1-P_2(\cos\theta)\right]-C,
\label{eq:quasipot}
\end{eqnarray}
where $\omega_A=\sqrt{B_0^2/(4\pi\rho_cR^2)}$ is the characteristic
Alfv\'en frequency of the star and $C$ is a constant. Here, $P_2(\cos\theta)=(3\cos^2\theta-1)/2$ denotes the Legendre polynominal of order $2$.

Since the potential $\Phi$ is the quantity of order of $GM/R$ with $M$ being the mass of the star,
the ratio of the second term on the right hand side of equation (\ref{eq:quasipot}) to $\Phi$
may be given by
$
\bar\omega_A^2\equiv{\omega_A^2/ \Omega_K^2}, 
$
where $\Omega_K=\sqrt{GM/R^3}$. 
For a neutron star model of the mass $M=1.4M_\odot$ and radius $R=10^6\rm cm$, for example,
we have $\bar\omega_A^2\simeq 2\times10^{-5}$ for the field strength $B_0=10^{16}$G, suggesting that
the effects of the magnetic field on the equilibrium structure is not significant so long as $B_0\la 10^{17}$G.
In this paper, as mentioned before, we assume that the magnetic field is sufficiently
weak so that the deformation of the equilibrium structure due to the magnetic field
can be treated as a small perturbation to 
the non-magnetic stars 
when $\bar\omega_A^2\ll 1$.
Under this assumption, we can regard $\hat{\rho}$ appearing in the terms proportional to $\omega_A^2$ in equation (\ref{eq:quasipot}) as
the density $\hat\rho_0$ in the non-magnetic star.
Thus, $\Psi$ satisfies
\begin{eqnarray}
\nabla^2\Psi=4\pi G\rho+\frac{1}{3}\omega_A^2\left[r^2{d^2\hat{\rho}_0\over dr^2}+6r{d\hat{\rho}_0\over dr}+6\hat{\rho}_0-\left(r^2{d^2\hat{\rho}_0\over dr^2}
+6r{d\hat{\rho}_0\over dr}\right)P_2(\cos\theta)\right]\,.
\label{eq:poisson_mod}
\end{eqnarray}
Since $\rho$ can be regarded as a function of $\Psi$ from equations (\ref{eq:hydro}) and (\ref{eq:poly}),
if we expand $\Psi (r,\theta)$ as
\begin{eqnarray}
\Psi (r,\theta)=\Psi_0(r)-2R^2\omega_A^2\left[\psi_0(x)+\psi_2(x)P_2(\cos\theta)\right],
\label{eq:psi_expand}
\end{eqnarray}
we may expand $\rho(r,\theta)$ as
\begin{eqnarray}
\rho (r,\theta)=\rho_0(r)-2R^2\omega_A^2\dfn{\rho_0}{\Psi_0}{}\left[\psi_0(x)+\psi_2(x)P_2(\cos\theta)\right].
\label{eq:rho_expand}
\end{eqnarray}
Here, $\Psi_0(r)=\Phi_0(r)$ and $\rho_0(r)$ are the gravitational potential and the density of the
non-magnetized star, and they satisfy $dp_0/dr=-\rho_0d\Phi_0/dr=-\rho_0GM_r/r^2$, $M_r=\int_0^r4\pi r^2\rho_0dr$, and $p_0=K_c\rho_0^{1+1/n}$.

Substituting equations (\ref{eq:psi_expand}) and (\ref{eq:rho_expand}) into (\ref{eq:poisson_mod}), we find 
\begin{eqnarray}
R^2\nabla^2\left[\psi_0(x)+\psi_2(x)P_2(\cos\theta)\right]=4\pi GR^2\dfn{\rho_0}{\Psi_0}{}\left[\psi_0(x)+\psi_2(x)P_2(\cos\theta)\right] +f_0(x)+f_2(x)P_2(\cos\theta), \end{eqnarray}
from which we obtain the following linear ordinary differential equations
for $\psi_0(x)$ and $\psi_2(x)$:
\begin{eqnarray}
\frac{1}{x^2}\dfn{}{x}{}\left(x^2\dfn{\psi_0}{x}{}\right)=k(x)\psi_0+f_0(x),
\label{eq:diffpsi0}
\end{eqnarray}
\begin{eqnarray}
\frac{1}{x^2}\dfn{}{x}{}\left(x^2\dfn{\psi_2}{x}{}\right)=\left[k(x)+\frac{6}{x^2}\right]\psi_2
+f_2(x),
\label{eq:diffpsi2}
\end{eqnarray}
where
\be
f_0(x)=-\frac{1}{6}\left(r^2{d^2\hat{\rho}_0\over dr^2}+6r{d\hat{\rho}_0\over dr}+6\hat{\rho}_0\right), \quad
f_2(x)=\frac{1}{6}\left(r^2{d^2\hat{\rho}_0\over dr^2}+6r{d\hat{\rho}_0\over dr}\right),
\ee
and
\begin{eqnarray}
k(x)=4\pi GR^2\dfn{\rho_0}{\Psi_0}{}.
\end{eqnarray}
In order to numerically integrate the differential equations (\ref{eq:diffpsi0}) and (\ref{eq:diffpsi2})
from the stellar center, we need to impose the regularity condition at the center, which may be obtained by substituting the expansion around the center $x=0$
\be
\psi_j=x^s\sum_{n=0}^\infty\left(a_n^{(j)}x^n\right) 
\ee
into (\ref{eq:diffpsi0}) and (\ref{eq:diffpsi2}) for $j=0$ and $j=2$, respectively.
Since $k(x)\rightarrow k(0)$, $f_0(x)\rightarrow f_0(0)$, 
and $f_2(x)\rightarrow f_{20}x^2$ as $x\rightarrow 0$, where $k(0)$, $f_0(0)$, and $f_{20}$ are constants, values of the exponent $s$ are given by $s=j$ 
for the regular solution at the center and 
the expansion coefficients $a^{(0)}_0$ and $a^{(2)}_0$ remain undetermined. Assuming that the density at the center $x=0$ is independent of $\omega_A$, we have 
$a^{(0)}_0=0$ for $\psi_0$. The expansion coefficient $a^{(2)}_0$ for $\psi_2$ must be specified by applying the surface boundary condition:
\be
3\psi_2(1)+{d\psi_2\over dx}(1)={1\over 6}{d\hat\rho\over dx}(1).
\ee
See Appendix~B for the derivation of the boundary condition.

\section{Perturbation equations}

For the modal analysis of magnetically deformed stars, we introduce the parameter $a$ that labels equi-potential surfaces of $\Psi(r,\theta)$. The parameter $a$ is defined such that
$\Psi(r,\theta)=\Psi_0(a)$, that is,
\begin{eqnarray}
\Psi_0(a)=\Psi_0(r)-2R^2\omega_A^2\left[\psi_0(x)+\psi_2(x)P_2(\cos\theta)\right],
\label{eq:equip}
\end{eqnarray}
which may define the equipotential surface as given by a function $r(a,\theta)$.
Assuming the deviation of the equipotential surface $r=r(a,\theta)$ from the spherical surface $r=a$ is 
small, we define the function $r(a,\theta)$ as
\begin{eqnarray}
r=a\left[1+\epsilon (a,\theta)\right],
\label{eq:raeps}
\end{eqnarray}
and we assume that $\epsilon$ is the quantity of order of $R^2\omega_A^2/\Psi_0(R)$. 
By substituting
equation (\ref{eq:raeps}) into (\ref{eq:equip}), we obtain the explicit expression for the function $\epsilon (a,\theta)$ up to the order of $\omega_A^2$:
\begin{eqnarray}
\epsilon (a,\theta)=\alpha (a)+\beta(a)P_2(\cos\theta),
\end{eqnarray}
where
\begin{eqnarray}
\alpha(a)=\frac{2c_1\bar{\omega}_A^2}{x^2}\psi_0(x),\quad \beta(a)=\frac{2c_1\bar{\omega}_A^2}{x^2}\psi_2(x),
\end{eqnarray}
where $c_1=(a/R)^3/[M(a)/M]$, 
and $M(a)$ denotes the mass inside the $a$-constant surface and $M=M(R)$.

Hereafter, we employ the parameter $a$ instead of the polar radial coordinate $r$ as the radial coordinate. In this coordinate system $(a,\theta,\phi)$, 
the line element is given by
\begin{eqnarray}
\rd s^2=(1+2\epsilon)(\rd a^2+a^2\rd\theta^2+a^2\sin^2\theta\rd\phi^2)+2a\pdn{\epsilon}{a}{}\rd a^2+2a\pdn{\epsilon}{\theta}{}\rd a\rd\theta.
\end{eqnarray}
Note that in this coordinate system, the pressure, the density and the effective potential of a magnetized star depend only on the radial coordinate $a$, although the orthogonality of the basis vectors is lost.

The governing equations of non-radial oscillations of a magnetized and uniformly 
rotating star are obtained by linearizing the basic equations. 
As for rotation effects on the oscillations, as mentioned in the previous section,  
we consider only the effects of the Coriolis force and ignore
those of the centrifugal force, where we assume the rotation axis is parallel to the magnetic axis.
Since the equilibrium state is assumed to be stationary
and axisymmetric, the perturbation quantities are proportional to
$\exp(i\omega t+im\phi)$, where $\omega$ is the frequency observed in
an inertial frame and $m$ is the azimuthal wave number. 
Then, the
linearized basic equations which govern the adiabatic, non-radial
oscillations of a magnetized and uniformly rotating star are written in the coordinate system $(a,\theta,\phi)$, to second order in $\omega_A$, as 
(Saio 1981; Lee 1993; Yoshida \& Lee 2000a)
\begin{eqnarray}
-\sigma^2\left[(1+2\epsilon)\pmb{\xi}+a\xi^a\nabla_0\epsilon+a(\pmb{\xi}\cdot\nabla_0\epsilon)\pmb{e}_a\right]=-\nabla_0\Phi'-\frac{1}{\rho}\nabla_0p'
+\frac{\rho'}{\rho^2}\left[ \dfn{p}{a}{}\pmb{e}_a -{1\over 4\pi}\left(\nabla_0\times\pmb{B}\right)\times\pmb{B} \right]+i\sigma\pmb{D}
\nonumber\\
+\frac{1}{4\pi\rho}\left[(\nabla_0\times\pmb{B}')\times\pmb{B}+(\nabla_0\times\pmb{B})\times\pmb{B}'\right], 
\label{eq:momentumeq}
\end{eqnarray}
\begin{eqnarray}
\rho'+\nabla_0\cdot(\rho\pmb{\xi})+\rho\pmb{\xi}\cdot\nabla_0\left(3\epsilon+a\pdn{\epsilon}{a}{}\right)=0,
\label{eq:continuityeq}
\end{eqnarray}
\begin{eqnarray}
\frac{\rho'}{\rho}=\frac{p'}{\Gamma_1p}-\frac{\xi^a}{a}aA, 
\label{eq:rhoprime}
\end{eqnarray}
\begin{eqnarray}
\left(B'\right)^i=\frac{1}{\sqrt{g}}\epsilon^{ijk}\frac{\partial}{\partial x^j}\left(\sqrt{g}\epsilon_{lmk}\xi^lB^m\right),
\label{eq:inductioneq}
\end{eqnarray}
where $\sigma = \omega+m\Omega$ denotes the oscillation frequency observed in the corotating frame of the star, 
$\pmb{\xi}(a,\theta,\phi)$ is the displacement vector, the prime ($'$) indicates the Eulerian perturbation, $\epsilon_{ijk}$ and $\epsilon^{ijk}$ 
are the Levi-Civita permutation symbols, $g$ is the determinant of the
metric $g_{ij}$, and 
\begin{eqnarray}
\nabla_0=\lim_{\epsilon\rightarrow 0} \left[ \pmb{e}_a\pdn{}{a}{}+\pmb{e}_\theta\frac{1}{a}\pdn{}{\theta}{}+\pmb{e}_\phi\frac{1}{a\sin\theta}\pdn{}{\phi} {}  \right] ,
\end{eqnarray}
and $\pmb{e}_a$, $\pmb{e}_\theta$, and $\pmb{e}_\phi$ are the basis vectors in the $a$, $\theta$, and 
$\phi$ directions, respectively.
Here, 
the vector $\pmb{D}$ in equation (\ref{eq:momentumeq}) comes from the Coriolis
force and is given by (see, e.g., Lee 1993; Yoshida \& Lee 2000a)
\begin{eqnarray}
D_a=2\Omega\left(1+2\epsilon+a\pdn{\epsilon}{a}{}\right)\sin\theta\xi^\phi,
\quad
D_\theta=2\Omega\left(1+2\epsilon+\frac{\sin\theta}{\cos\theta}\pdn{\epsilon}{\theta}{}\right)\cos\theta\xi^\phi, \nonumber
\end{eqnarray}
\begin{eqnarray}
D_\phi=-2\Omega\left[\left(1+2\epsilon+a\pdn{\epsilon}{a}{}\right)\sin\theta\xi^a+\left(1+2\epsilon+\frac{\sin\theta}{\cos\theta}\pdn{\epsilon}{\theta}{}\right)\cos\theta\xi^\theta\right],
\end{eqnarray}
and $aA$ in equation (\ref{eq:rhoprime}) denotes the Schwarzschild discriminant defined as
\be
aA=\dfn{\ln\rho}{\ln a}{}-\frac{1}{\Gamma_1}\dfn{\ln p}{\ln a}{},
\ee
where $\Gamma_1=(\partial\ln p/\partial\ln\rho)_{\rm ad}$.
In this paper, for simplicity, we employ the Cowling approximation, neglecting $\Phi'$.

Because of the Lorentz and Coriolis terms in the equation of motion (\ref{eq:momentumeq}),
separation of variables for the perturbations is
impossible between the radial coordinate $(a)$ and the angular
coordinates $(\theta,\phi)$. We therefore expand the perturbations in
terms of the spherical harmonic functions $Y_l^m(\theta,\phi)$ with
different $l$'s for a given azimuthal index $m$. 
The displacement vector $\pmb{\xi}$ is then given by (see e.g., Lee 2005, 2007)
\begin{eqnarray}
\xi^a=\sum_{j=1}^{j_{\rm max}}aS_{l_j}(a)Y_{l_j}^m(\theta,\phi),
\label{eq:xiaexp}
\end{eqnarray}
\begin{eqnarray}
\xi^\theta=\sum_{j=1}^{j_{\rm max}}\left[aH_{l_j}(a)\pdn{}{\theta}{}Y_{l_j}^m(\theta,\phi)-iaT_{l_j'}(a)\frac{1}{\sin\theta}\pdn{}{\phi}{}Y_{l_j'}^m(\theta,\phi)\right],
\label{eq:xithetaexp}
\end{eqnarray}
\begin{eqnarray}
\xi^\phi=\sum_{j=1}^{j_{\rm max}}\left[aH_{l_j}(a)\frac{1}{\sin\theta}\pdn{}{\phi}{}Y_{l_j}^m(\theta,\phi)+iaT_{l_j'}(a)\pdn{}{\theta}{}Y_{l_j'}^m(\theta,\phi)\right],
\label{eq:xiphiexp}
\end{eqnarray}
and the vector $\pmb{B}'$ is given by
\begin{eqnarray}
\frac{(B^a)'}{k\rho}=\sum_{j=1}^{j_{\rm max}}iah_{l_j}^S(a)Y_{l_j}^m(\theta,\phi),
\label{eq:baexp}
\end{eqnarray}
\begin{eqnarray}
\frac{(B^\theta)'}{k\rho}=\sum_{j=1}^{j_{\rm max}}\left[iah_{l_j}^H(a)\pdn{}{\theta}{}Y_{l_j}^m(\theta,\phi)-ah_{l_j'}^T(a)\frac{1}{\sin\theta}\pdn{}{\phi}{}Y_{l_j'}^m(\theta,\phi)\right],
\label{eq:bthetaexp}
\end{eqnarray}
\begin{eqnarray}
\frac{(B^\phi)'}{k\rho}=\sum_{j=1}^{j_{\rm max}}\left[iah_{l_j}^H(a)\frac{1}{\sin\theta}\pdn{}{\phi}{}Y_{l_j}^m(\theta,\phi)+ah_{l_j'}^T(a)\pdn{}{\theta}{}Y_{l_j'}^m(\theta,\phi)\right],
\label{eq:bphiexp}
\end{eqnarray}
where $l_j=|m|+2(j-1)$ and $l_j'=l_j+1$ for even modes, and
$l_j=|m|+2j-1$ and $l_j'=l_j-1$ for odd modes, respectively, and
$j=1,2,3,...,j_{{\rm{max}}}$.

The Euler perturbations of the pressure and density are given by
\begin{eqnarray}
p'=\sum_{j=1}^{j_{\rm max}}p_{l_j}'(a)Y_{l_j}^m(\theta,\phi),\quad
\rho'=\sum_{j=1}^{j_{\rm max}}\rho_{l_j}'(a)Y_{l_j}^m(\theta,\phi).
\label{eq:prhoexp}
\end{eqnarray}
In this paper, we usually use
$j_{{\rm{max}}}=12$ to obtain solutions with sufficiently high-angular resolution.  
Substituting the expansions (\ref{eq:xiaexp})-
(\ref{eq:prhoexp})
into the linearized basic equations (\ref{eq:momentumeq})-(\ref{eq:inductioneq}), we obtain a finite set of coupled
linear ordinary differential equations for the expansion coefficients
$S_{l_j}(a)$ and $p^\prime_{l_j}(a)$, which we call the oscillation
equations to be solved in the interior of magnetized and uniformly rotating stars. The
set of oscillation equations obtained for the magnetized rotating star is given in Appendix~A. 
\begin{figure}
\begin{center}
\resizebox{0.39\columnwidth}{!}{
\includegraphics{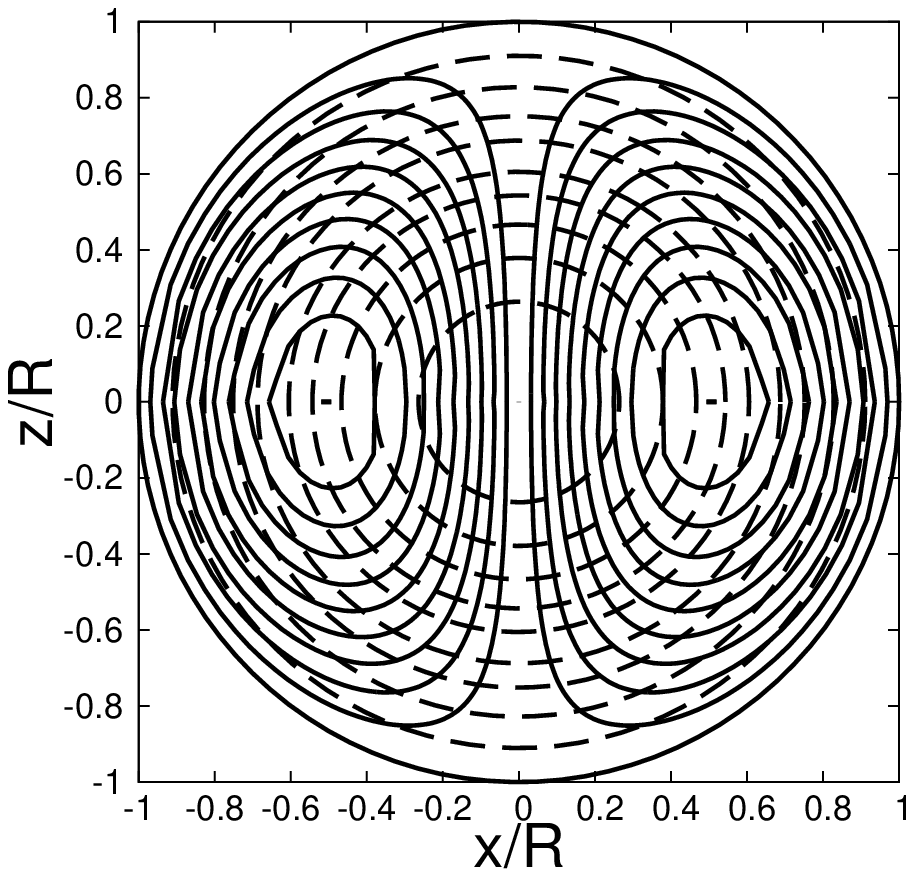}}
\hspace*{-1.75cm}
\resizebox{0.39\columnwidth}{!}{
\includegraphics{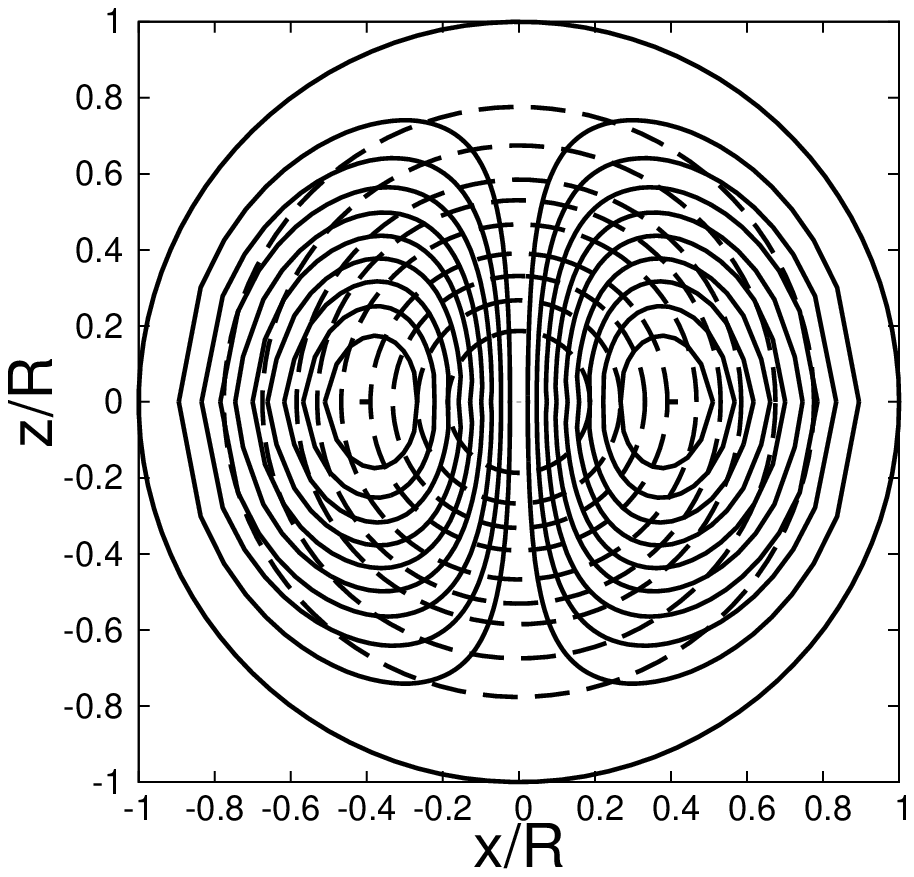}}
\hspace*{-1.75cm}
\resizebox{0.39\columnwidth}{!}{
\includegraphics{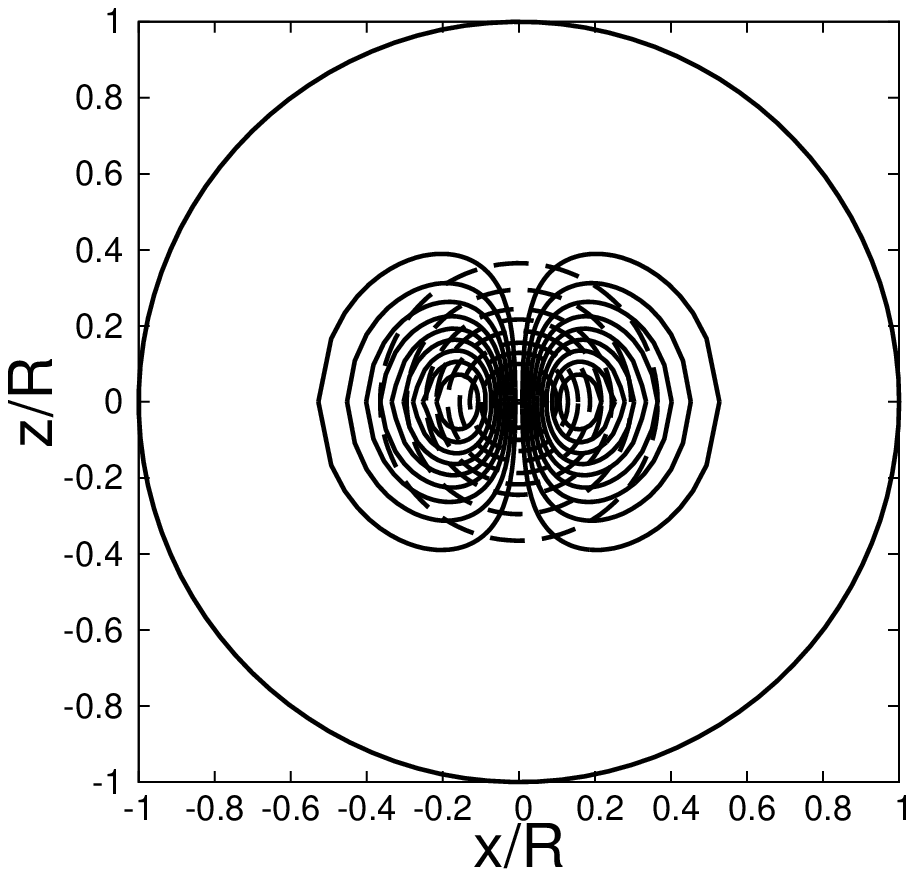}}
\end{center}
\caption{Equi-magnetic field strength  contours (solid lines) and equi-density
  contours (long-dashed lines) on the meridional cross sections are
  plotted, from left to right panels, for the polytropes of the index $n=$1, 1.5, and 3, respectively. 
  The outer-most solid circles show stellar surfaces. The solid contours correspond to $B/B_{\rm max}=0.1$, $0.2$, $0.3$, $0.4$, 
  $0.5$, $0.6$, $0.7$, $0.8$, and $0.9$, and the long-dashed contours to $\rho/\rho_c=0.1$, $0.2$, $0.3$, $0.4$, $0.5$, $0.6$,
  $0.7$, $0.8$, and $0.9$.}\label{contour_plots}
\end{figure}

\section{Numerical results}

\begin{figure}
\begin{center}
\resizebox{0.5\columnwidth}{!}{
\includegraphics{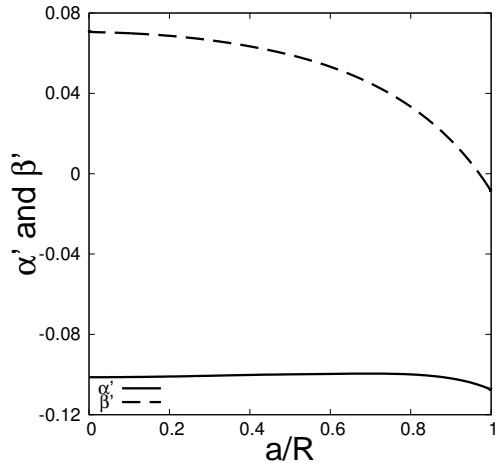}}
\end{center}
\caption{Functions $\alpha'\equiv\alpha/\bar\omega^2_A$ and $\beta'\equiv\beta/\bar\omega_A^2$ versus the fractional radius $a/R$ for the $n=1$ polytrope. 
}\label{alpha_beta}
\end{figure}

In this study, the three polytropes with indices $n=1$, $1.5$, and $3$ are used for modal analyses of the magnetized stars. The polytropes with $n=1$ and $1.5$ 
and with $n=3$ are regarded as simplified models of the neutron star and the normal star, respectively. For these polytopes, distributions of the density  
and the imposed magnetic fields are shown in Fig~\ref{contour_plots}. In this figure, the equi-density and equi-magnetic field strength contours on the meridional 
cross sections are given.  We see that the density and magnetic field distribution of the models with a larger polytropic index tend to be concentrated in the central 
region of the star. Typical values of the mass $M$ and radius $R$ for the neutron star and the normal star are assumed to be $(M, R)=(1.4M_\odot,10^6 {\rm cm})$ 
and $(M, R)=(M_\odot, R_\odot)$, respectively. Thus, we have $\bar\omega_A=4.42\times 10^{-3}$ 
for the neutron 
star with the field strength $B_0=10^{16}$G, and $\bar\omega_A=7.39\times10^{-4}$ for the normal star with the field strength $B_0=10^6$G.
For the polytrope with $n=1$, the functions $\alpha(a)/\bar\omega_A^2$ and $\beta(a)/\bar\omega_A^2$ for the magnetic deformation are plotted as functions of 
$a/R$ in Fig.~\ref{alpha_beta}.

\subsection{$g$-, $f$-, and $p$-modes}

We first calculate $f$-, low radial order $g$-,  and $p$-modes of the polytropes taking account of the effects of the toroidal magnetic field. In these calculations, 
no effects of rotation are considered. To study oscillation modes for the neutron star and normal star models, the adiabatic indices for perturbations are 
assumed to be 
\be
{1\over\Gamma_1}={n\over n+1}+\gamma
\ee
with $\gamma$ being a constant, for which
$
aA=\gamma{(d\ln p/ d\ln a)}.
$
In this subsection, we use $\gamma=-10^{-4}$ for the polytropes with the indices $n=1$ and 1.5, but for the polytrope
with $n=3$, we assume $\Gamma_1=5/3$, and hence $\gamma=-3/20$.
Since all the magnetic terms in the oscillation equations are proportional to $\bar\omega_A^2$, an oscillation frequency of the modes may be written by 
(see Appendices~A \& C and Unno et al. 1989)
\begin{eqnarray}
\bar\sigma
= \bar\sigma_0+\bar E_2\bar{\omega}_A^2+\cdots,
\end{eqnarray}
where $\bar\sigma_0$
is the oscillation frequency of the non-magnetized star, and
$\bar E_2$ is a proportionality coefficient and can be obtained by calculating
the oscillation frequency of the mode for two different values of $\bar\omega_A^2$, that is, 
$\bar\omega_A^2=0$ and $\sim 10^{-6}$, for example.
Here, $\bar\sigma_0$ and $\bar E_2$ are the quantities normalized by the Kepler frequency of the star $\Omega_k$.
This coefficient $\bar E_2$ for a mode may also be calculated by using the eigenfunctions of the non-magnetized
star by treating $\bar\omega_A^2$ as a small parameter.
We have used the symbol $\bar E_2^\prime$ to denote the coefficient computed by using the eigenfunctions for the non-magnetized 
star. The derivation and the explicit expression for $\bar E_2^\prime$ are given in Appendix~C. 
\begin{table*}
\begin{center}
\caption{Coefficients $E_2$, $E_2^\prime$, and $\bar E_2^0$ for $g$-, $f$-, and $p$-modes of $l=m$ for 
the polytropic model with $n=1$ and $\gamma=-10^{-4}$ $^{*}$}
\label{symbols}
\begin{threeparttable}
\begin{tabular}{@{}crrrr}
\hline
\hline
mode & $ \ \bar\sigma_0$ & $ \ \bar E_2$ & $ \ \bar E_2^\prime$ & $ \ \bar E_2^0$ \\
\hline
&& $m=1$ && \\
\hline
$g_3\cdots\cdots$ & 0.00570 & -1.578($+0$) & -1.490($+0$) & -1.491($+0$) \\
$g_2\cdots\cdots$ & 0.00770 & -2.051($+0$) & -2.095($+0$) & -2.095($+0$) \\
$g_1\cdots\cdots$ & 0.01203 & -3.003($+0$) & -3.063($+0$) & -3.064($+0$) \\
$p_1\cdots\cdots$ & 3.26931 & 6.162($-1$) & 6.164($-1$) & 3.553($-1$) \\
$p_2\cdots\cdots$ & 5.09325 & 1.113($+0$) & 1.114($+0$) & 5.650($-1$) \\
$p_3\cdots\cdots$ & 6.85013 & 1.550($+0$) & 1.554($+0$) & 7.586($-1$) \\
\hline
 && $m=2$ && \\
\hline
$g_3\cdots\cdots$ & 0.00884 & 3.518($+1$) & 3.488($+1$) & 3.487($+1$) \\
$g_2\cdots\cdots$ & 0.01152 & 2.427($+1$) & 2.402($+1$) & 2.402($+1$) \\
$g_1\cdots\cdots$ & 0.01678 & 1.133($+1$) & 1.130($+1$) & 1.130($+1$) \\
$f \ \cdots\cdots$ & 1.65562 & 4.107($-1$) & 4.109($-1$) & 3.567($-1$) \\
$p_1\cdots\cdots$ & 3.79225 & 7.099($-1$) & 7.103($-1$) & 4.387($-1$) \\
$p_2\cdots\cdots$ & 5.67886 & 1.173($+0$) & 1.175($+0$) & 6.713($-1$) \\
$p_3\cdots\cdots$ & 7.48089 & 1.605($+0$) & 1.608($+0$) & 8.874($-1$) \\
\hline
&& $m=3$ && \\
\hline
$g_3\cdots\cdots$ & 0.01136 & 7.583($+1$) & 7.498($+1$) & 7.498($+1$) \\
$g_2\cdots\cdots$ & 0.01443 & 5.352($+1$) & 5.319($+1$) & 5.319($+1$) \\
$g_1\cdots\cdots$ & 0.02008 & 2.732($+1$) & 2.732($+1$) & 2.731($+1$) \\
$f \ \cdots\cdots$ & 1.97094 & 6.696($-1$) & 6.696($-1$) & 5.545($-1$) \\
$p_1\cdots\cdots$ & 4.22956 & 8.280($-1$) & 8.282($-1$) & 5.196($-1$) \\
$p_2\cdots\cdots$ & 6.18783 & 1.272($+0$) & 1.272($+0$) & 7.593($-1$) \\
$p_3\cdots\cdots$ & 8.04091 & 1.701($+0$) & 1.703($+0$) & 9.900($-1$) \\
\hline
\end{tabular}
\medskip
\begin{tablenotes}
\item[*] We use the notation of $1.000\times 10^N \equiv 1.000(N)$.
\end{tablenotes}
\end{threeparttable}
\end{center}
\end{table*}
\begin{table*}
\begin{center}
\caption{Coefficients $E_2$, $E_2^\prime$, and $\bar E_2^0$ for $g$-, $f$-, and $p$-modes of $l=m$ for 
the polytropic model with $n=1.5$ and $\gamma=-10^{-4}$ $^{*}$}
\label{symbols}
\begin{threeparttable}
\begin{tabular}{@{}crrrr}
\hline
\hline
mode & $\ \bar\sigma_0$ & $\ \bar E_2$ & $\ \bar E_2^\prime$ & $\ \bar
E_2^0$ \\
\hline
&& $m=1$ && \\
\hline
$g_3\cdots\cdots$ & 0.00788 & -8.899($-1$) & -8.120($-1$) & -8.123($-1$) \\
$g_2\cdots\cdots$ & 0.01057 & -1.179($+0$) & -1.193($+0$) & -1.193($+0$) \\
$g_1\cdots\cdots$ & 0.01626 & -1.797($+0$) & -1.856($+0$) & -1.856($+0$) \\
$p_1\cdots\cdots$ & 3.08199 & 5.763($-1$) & 5.761($-1$) & 9.406($-2$) \\
$p_2\cdots\cdots$ & 4.64233 & 9.672($-1$) & 9.668($-1$) & 1.600($-1$) \\
$p_3\cdots\cdots$ & 6.15692 & 1.329($+0$) & 1.328($+0$) & 2.190($-1$) \\
\hline
 && $m=2$ && \\
\hline
$g_3\cdots\cdots$ & 0.01214 & 2.316($+1$) & 2.286($+1$) & 2.286($+1$) \\
$g_2\cdots\cdots$ & 0.01564 & 1.628($+1$) & 1.607($+1$) & 1.607($+1$) \\
$g_1\cdots\cdots$ & 0.02217 & 8.198($+0$) & 8.153($+0$) & 8.152($+0$) \\
$f \ \cdots\cdots$ & 1.84930 & 4.027($-1$) & 4.027($-1$) & 1.621($-1$) \\
$p_1\cdots\cdots$ & 3.55537 & 7.093($-1$) & 7.093($-1$) & 1.205($-1$) \\
$p_2\cdots\cdots$ & 5.14850 & 1.073($+0$) & 1.073($+0$) & 1.808($-1$) \\
$p_3\cdots\cdots$ & 6.69114 & 1.429($+0$) & 1.428($+0$) & 2.454($-1$) \\
\hline
&& $m=3$ && \\
\hline
$g_3\cdots\cdots$ & 0.01547 & 4.950($+1$) & 4.872($+1$) & 4.872($+1$) \\
$g_2\cdots\cdots$ & 0.01935 & 3.546($+1$) & 3.511($+1$) & 3.511($+1$) \\
$g_1\cdots\cdots$ & 0.02596 & 1.941($+1$) & 1.938($+1$) & 1.938($+1$) \\
$f \ \cdots\cdots$ & 2.15084 & 5.640($-1$) & 5.640($-1$) & 2.301($-1$) \\
$p_1\cdots\cdots$ & 3.93952 & 8.393($-1$) & 8.391($-1$) & 1.504($-1$) \\
$p_2\cdots\cdots$ & 5.58066 & 1.192($+0$) & 1.191($+0$) & 2.007($-1$) \\
$p_3\cdots\cdots$ & 7.15896 & 1.544($+0$) & 1.542($+0$) & 2.653($-1$) \\
\hline
\end{tabular}
\medskip
\begin{tablenotes}
\item[*] We use the notation of $1.000\times 10^N \equiv 1.000(N)$.
\end{tablenotes}
\end{threeparttable}
\end{center}
\end{table*}
\begin{table*}
\begin{center}
\caption{Coefficients $E_2$, $E_2^\prime$, and $\bar E_2^0$ for $g$-, $f$-, and $p$-modes of $l=m$ for 
the polytropic model with $n=3$ and $\Gamma_1=5/3$ $^{*}$}
\label{symbols}
\begin{threeparttable}
\begin{tabular}{@{}crrrr}
\hline
\hline
mode & $\ \bar\sigma_0$ & $\ \bar E_2$ & $\ \bar E_2^\prime$ & $\ \bar
E_2^0$ \\
\hline
&& $m=1$ && \\
\hline
$g_3\cdots\cdots$ & 0.88994 & -1.088($-3$) & -7.875($-4$) & -7.747($-3$) \\
$g_2\cdots\cdots$ & 1.16154 & -1.270($-3$) & -1.079($-3$) & -9.976($-3$) \\
$g_1\cdots\cdots$ & 1.68082 & 9.069($-4$) & 7.333($-4$) & -1.306($-2$) \\
$p_1\cdots\cdots$ & 3.81006 & 2.906($-1$) & 2.905($-1$) & 2.689($-3$) \\
$p_2\cdots\cdots$ & 5.01208 & 4.335($-1$) & 4.331($-1$) & 2.803($-3$) \\
$p_3\cdots\cdots$ & 6.25522 & 5.541($-1$) & 5.538($-1$) & 3.836($-3$) \\
\hline
 && $m=2$ && \\
\hline
$g_3\cdots\cdots$ & 1.35792 & 1.948($-1$) & 1.930($-1$) & 1.802($-1$) \\
$g_2\cdots\cdots$ & 1.70580 & 1.502($-1$) & 1.494($-1$) & 1.332($-1$) \\
$g_1\cdots\cdots$ & 2.29614 & 1.103($-1$) & 1.099($-1$) & 8.008($-2$) \\
$f \ \cdots\cdots$ & 3.06379 & 2.193($-1$) & 2.194($-1$) & 1.814($-2$) \\
$p_1\cdots\cdots$ & 4.14666 & 3.521($-1$) & 3.521($-1$) & 1.064($-2$) \\
$p_2\cdots\cdots$ & 5.39097 & 4.792($-1$) & 4.788($-1$) & 7.335($-3$) \\
$p_3\cdots\cdots$ & 6.65382 & 5.990($-1$) & 5.978($-1$) & 6.608($-3$) \\
\hline
&& $m=3$ && \\
\hline
$g_3\cdots\cdots$ & 1.70370 & 4.161($-1$) & 4.122($-1$) & 3.942($-1$) \\
$g_2\cdots\cdots$ & 2.07374 & 3.245($-1$) & 3.227($-1$) & 3.007($-1$) \\
$g_1\cdots\cdots$ & 2.64527 & 2.338($-1$) & 2.334($-1$) & 1.920($-1$) \\
$f \ \cdots\cdots$ & 3.12498 & 2.567($-1$) & 2.566($-1$) & 3.173($-2$) \\
$p_1\cdots\cdots$ & 4.37567 & 3.947($-1$) & 3.946($-1$) & 1.118($-2$) \\
$p_2\cdots\cdots$ & 5.68481 & 5.193($-1$) & 5.189($-1$) & 8.675($-3$) \\
$p_3\cdots\cdots$ & 6.98164 & 6.392($-1$) & 6.383($-1$) & 8.076($-3$) \\
\hline
\end{tabular}
\medskip
\begin{tablenotes}
\item[*] We use the notation of $1.000\times 10^N \equiv 1.000(N)$. 
\end{tablenotes}
\end{threeparttable}
\end{center}
\end{table*}

In Tables 1--3, we tabulate the coefficients $\bar E_2$ and $\bar E_2^\prime$ as well as the frequency $\bar\sigma_0$ for the $f$-modes and
low radial order $p$- and $g$-modes of $l=m=1$, 2, and 3 for the polytropes with the indices $n=1$, $1.5$, and $3$. We observe that the two 
coefficients $\bar E_2$ and $\bar E_2^\prime$ are in good agreement with each other, except for a few very low frequency $g$-modes. 
In these tables, we also tabulate $\bar E_2^0$, which is the same as the coefficient $\bar E_2^\prime$ but calculated ignoring all the equilibrium 
deformation effects due to magnetic stress. We see that the frequencies of the $f$- and $p$-modes are strongly affected by the equilibrium deformation, 
but the deformation effects are not very important for the $g$-modes. This property of the frequency responses to the deformation due to magnetic 
field is quite similar to that found for the rotational deformation
(Saio 1981). We note that the frequency we obtain for $f$ mode is
consistent with that by Lander et al. (2010) for
$\Omega/\sqrt{G\rho_c}\la$ 0.1 (because they consider the effects
of the second order of $\Omega$).

From Tables 1 and 2, we see modal properties of $f$-, low radial order $g$-,  and $p$-modes for the neutron star models. 
Because of the small value of $|\gamma|$, the frequencies $\bar\sigma_0$ of the $g$-modes are quite low, which 
may be consistent with almost isentropic stratifications expected in the deep interior of cold neutron stars.
We find the ratio $E_2/\sigma_0$ for the $g$-modes is much larger than the ratio for the $f$- and $p$-modes,
suggesting the low frequency $g$-modes are more susceptible to the magnetic field, reflecting their very low frequencies of order of 
$\sqrt{|\gamma|}$. The ratio $E_2/\sigma_0$ for the $g$-modes increases with $m$, while the ratio for $f$- and $p$-modes
only weakly depends on $m$.
From Table 3, we see modal properties of $f$-, low radial order $g$-,  and $p$-modes for the normal star model. 
The ratio $E_2/\sigma_0$ have almost the same order of magnitudes for the $g$-, $f$-, and $p$-modes, except for
the $m=1$ $g$-modes, for which the ratio is much smaller.
It is also interesting to note that the magnitudes of the ratio for the $g$-modes is of order of 0.1 (except for the $m=1$ $g$-modes), 
the value of which is much smaller than those for the $g$-modes of the polytropes of $n=1$ and 1.5 with $\gamma=-10^{-4}$.

The coefficient $\bar E_2$ in the tables 1 to 3 is computed by using two data points with different values of $\bar\omega_A^2$.
For example, 
$\bar E_2$ computed with 4 data points for $g_3$ modes of $m=1$ and $m=2$ for $n=1$ are $-1.577$ and $35.18$,
and the coefficients for $p_1$ modes of $m=1$ and $m=2$ are $0.6164$ and $0.7100$.
We therefore think that the coefficients $\bar E_2$ in the tables have at least three significant digits.

In Figure 3, we plot the expansion coefficients $S_{l_1}$, $H_{l_1}$, and $T_{l^\prime_1}$ of the $g_1$, $f$, and $p_1$-modes of $l=m=2$ for 
the $n=1$ polytrope with $B_0=10^{16}$ G. 
The first expansion coefficients associated with the harmonic degree $l_1$ and $l^\prime_1$
are dominant over the coefficients with $l_j$ and $l^\prime_j$ for $j>1$, and 
the difference in the dominant expansion coefficients of the modes between the magnetized and non-magnetized models
is almost indiscernible.
Because of the surface boundary condition (\ref{eq:surfaceboundary}) and an algebraic relation (\ref{eq:moment_theta}) in Appendix~A, 
when $\bar\Omega=0$ and $\bar\omega_A^2\ll 1$ we have
$H_{l_1}\simeq S_{l_1}/\bar\sigma^2$ at the surface and hence $H_{l_1}$ at the surface can be very large
for $g$-modes having very low frequencies $\bar\sigma\ll 1$ for the
normalization $S_{l_1}=1$. In Figure 4, we plot magnetic field perturbations
$S_{bl}$, $H_{bl}$, and $T_{bl'}$ of the $g_1$, $f$, and $p_1$-modes
of $l=m=2$ for the $n=1$ polytrope with $B_0=10^{16}$ G, where 
$S_{bl}\equiv k\rho ah_{l}^S/B_0$,
$H_{bl}\equiv k\rho ah_{l}^H/B_0$, and $T_{bl'}\equiv k\rho
ah_{l'}^T/B_0$.

\begin{figure}
\begin{center}
\resizebox{0.39\columnwidth}{!}{
\includegraphics{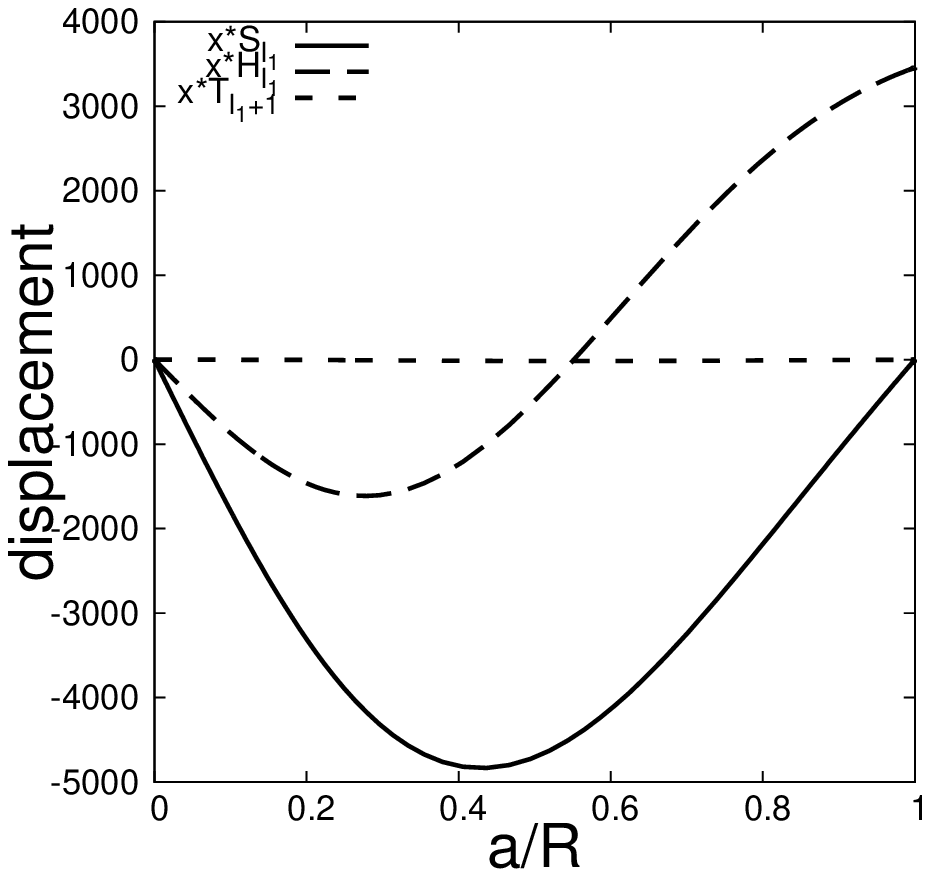}}
\hspace*{-1.75cm}
\resizebox{0.39\columnwidth}{!}{
\includegraphics{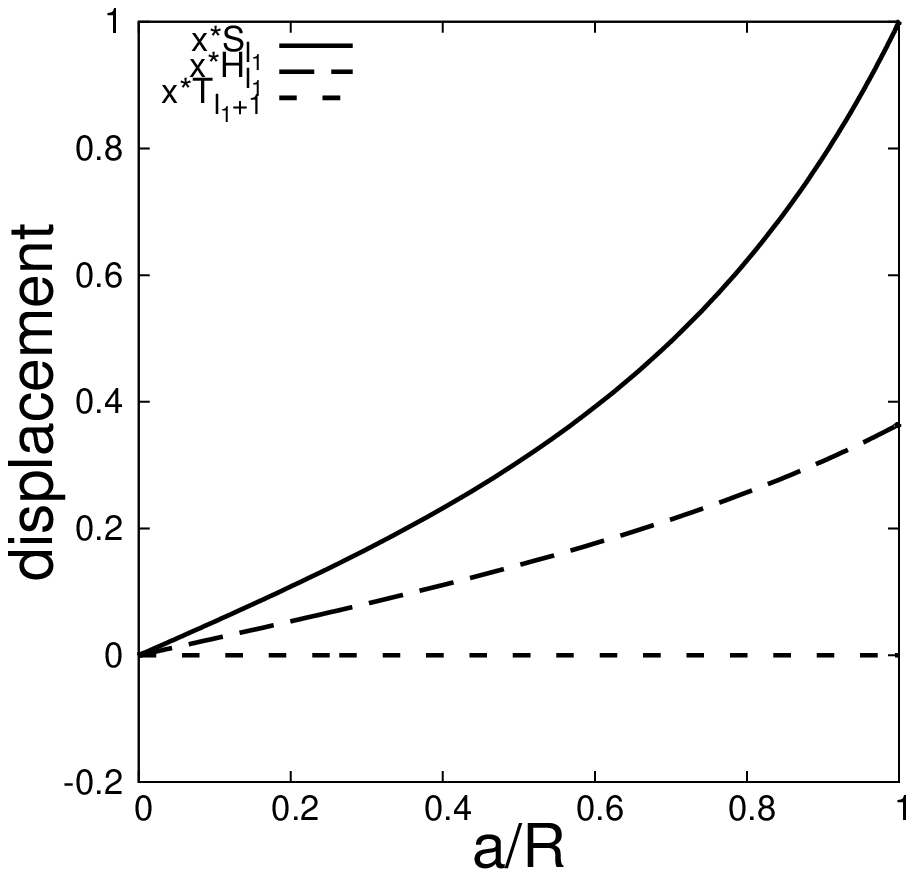}}
\hspace*{-1.75cm}
\resizebox{0.39\columnwidth}{!}{
\includegraphics{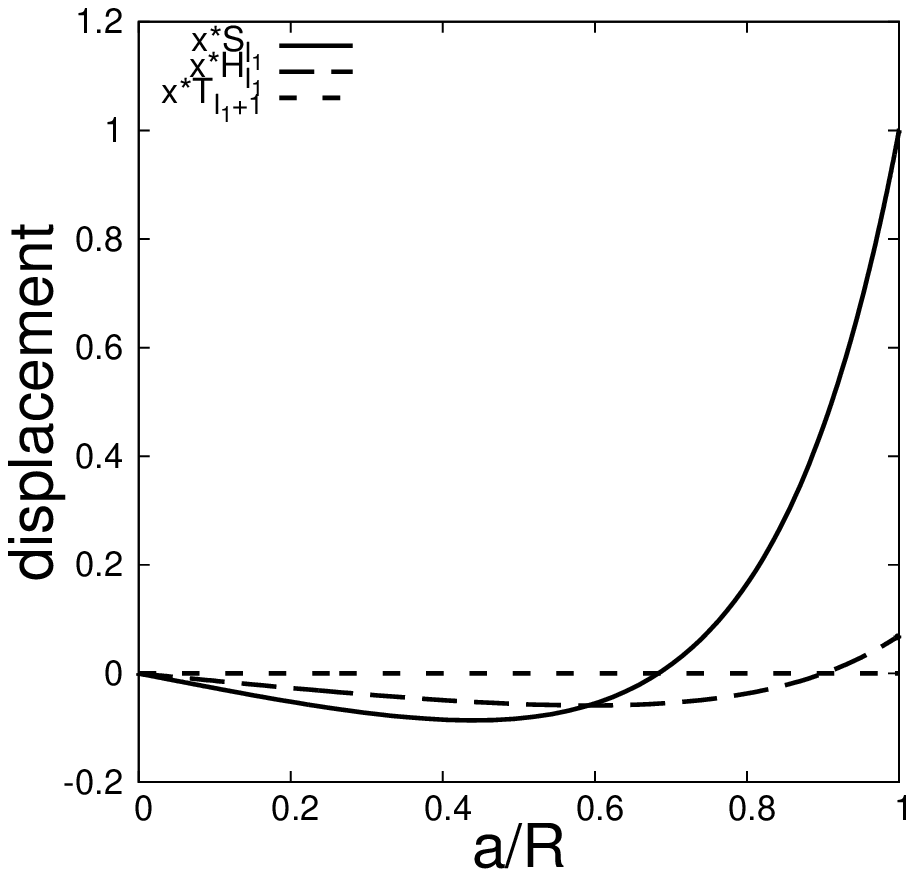}}
\end{center}
\caption{Eigenfunctions of $m=2$ even modes for the polytrope with $n=1$ and $\gamma=-10^{-4}$ for $B_0=10^{16}$
  G, where, from left to right panels, the eigenfunctions plotted are those of the $g_1$, $f$, and $p_1$ modes. 
  The solid lines, the long dashed lines and the short dashed lines 
  are for the functions $xS_{l_1}$,
  $xH_{l_1}$, and $xT_{l_1+1}$ with $x=a/R$, and the amplitude normalization is given by $S_{l_1}=1$ at the surface.}
\end{figure}

\begin{figure}
\begin{center}
\resizebox{0.39\columnwidth}{!}{
\includegraphics{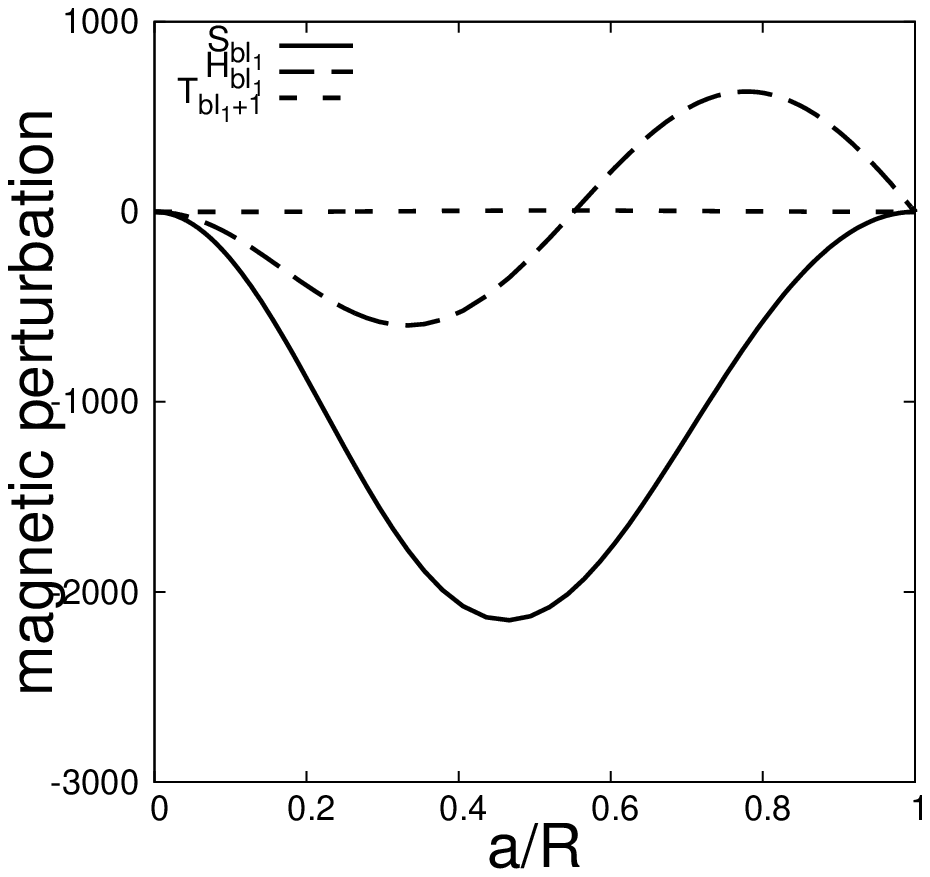}}
\hspace*{-1.69cm}
\resizebox{0.39\columnwidth}{!}{
\includegraphics{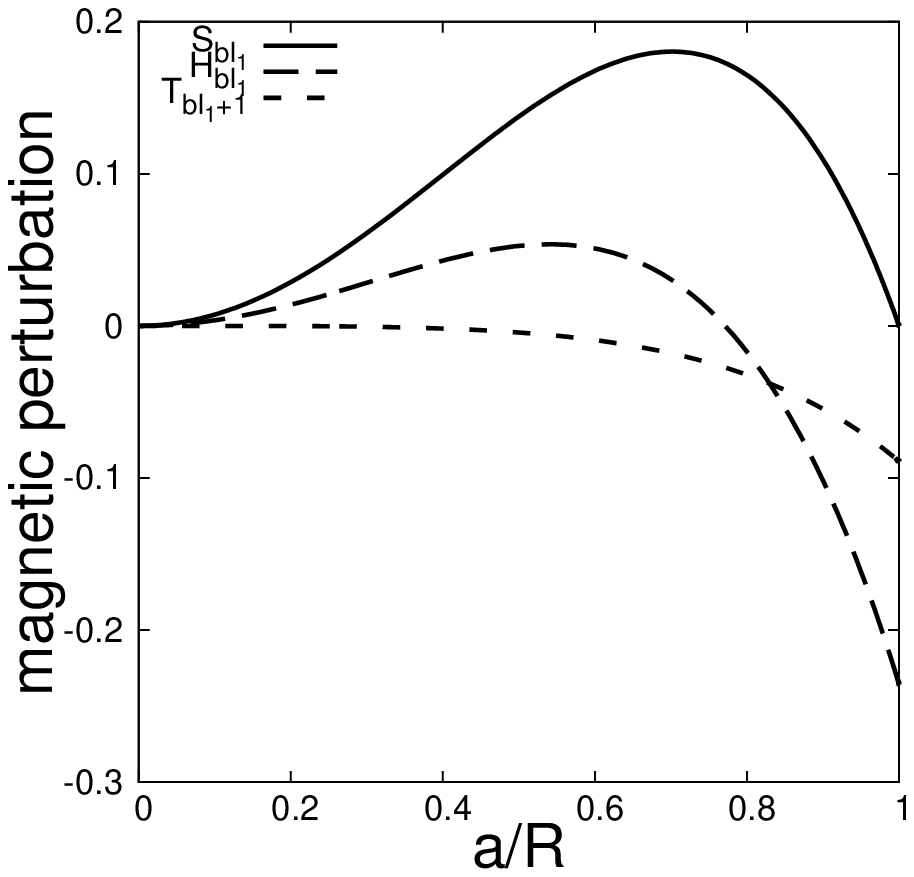}}
\hspace*{-1.69cm}
\resizebox{0.39\columnwidth}{!}{
\includegraphics{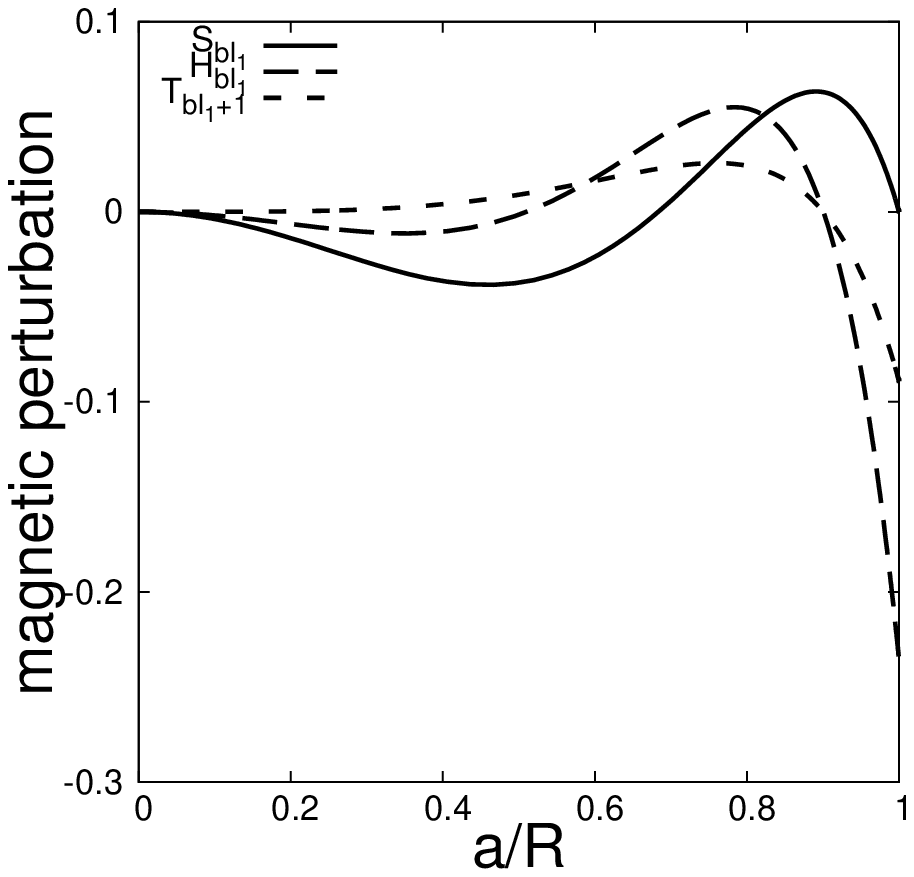}}
\end{center}
\caption{Same as Figure 3 but for the eigenfunctions
$S_{bl_1}\equiv k\rho a h_{l}^S/B_0$ (solid lines), $H_{bl_1}\equiv k\rho ah_{l}^H/B_0$ (long dashed lines), and 
$T_{bl_1+1}\equiv k\rho a  h_{l'}^T/B_0$ (short dashed lines).
}
\end{figure}

For slowly rotating stars, we may write the inertial frame oscillation
frequency $\omega$ of a mode as
\be
\omega=\omega_0+m(C_1-1)\Omega+E_2\bar\omega_A^2+\cdots,
\ee
where $C_1$ represents the response of the mode frequency to the slow rotation.
Since $\bar\omega_A^2\simeq 10^{-5}\sim 10^{-3}$ for $B_0=10^{16}\sim 10^{17}$G and $E_2\sim 10$,
the rotational effects may dominate the magnetic ones for
$\bar\Omega\ga 10^{-1}$.

\subsection{rotational modes}

We consider the effects of the magnetic field on rotational modes such as inertial modes and
$r$-modes, for which the Coriolis force is the restoring force and 
the oscillation frequency is proportional to the rotation frequency $\Omega$. 
As shown by Yoshida \& Lee (2000b), the stratification of the stellar interior strongly affects modal properties of the rotational mode. 
Since we are concerned with purely magnetic effects on the rotational mode, in this subsection, we focus on non-stratified stars or 
isentropic stars, in which the adiabatic index for perturbations is given by $\Gamma_1=1+n^{-1}$. 
To represent the effects of the magnetic field on the rotational modes,
it is convenient to use the frequency ratio $\kappa\equiv\sigma/\Omega$, where
$\sigma=\omega+m\Omega$ denotes the frequency observed in
the co-rotating frame of the star, and for small values of $\bar\omega_A^2(\ll\bar\Omega^2)$ we may write
\be
\kappa=\kappa_0(\Omega)\left[1+\eta_2{\bar\omega_A^2\over\bar\Omega^2}\right]+\cdots,
\ee
where the coefficient $\kappa_0$ may depend on the rotation rate
$\Omega$ and the coefficient $\eta_2$ is a constant in the limit of $\bar\omega_A^2/\bar\Omega^2\to 0$ (see  Appendix~C).
Inertial modes and $r$-modes of uniformly rotating isentropic polytropes were studied, for example, by Lockitch \& Friedman (1999) and Yoshida \& Lee (2000a).
The $r$-modes of $m\not=0$ and $l^\prime\ge |m|$ are non-axisymmetric and retrograde modes and the frequency ratio $\kappa_0$ tends to $2m/[l^\prime(l^\prime+1)]$
as $\Omega\rightarrow 0$.
The ratio $\kappa_0$ for an inertial mode also tends to a definite value as $\Omega\rightarrow 0$,
depending on $m$, $l$, and the polytropic index $n$ (see, e.g., Yoshida \& Lee 2000a), and
we may use $\kappa_0(0)$ as a labeling of the inertial modes for a given $m$.

Since stars with strong magnetic fields are frequently very slow rotators, 
we consider rotational modes in slowly rotating stars.
In Table 4, the coefficients $\eta_2$ and $\eta_2^\prime$ as
well as $\kappa_0$ are tabulated for the $l^\prime=|m|$ $r$-modes
and inertial modes for $m=2$ of isentropic (i.e., $\gamma=0$) polytropes with three different indices $n$. 
We use the symbol
$\eta_2^\prime$ to denote the coefficient computed by using
the eigenfunctions of non-magnetized slowly rotating stars. The
coefficient $\eta_2$ of the fitting formula $y=\eta_2x$, where
$x\equiv 1/\bar\Omega^2$ and $y\equiv E_2/\sigma_0$, can be calculated by
a least-square method.
In the table, $l_0-m=1$ means the
$r$-modes, and $l_0-m\ge 2$ correspond to inertial modes (see
Yoshida \& Lee 2000a), and the even and odd numbers of $l_0-m$ indicate even
and odd parity, respectively.  
Since using $\kappa_0$, the frequencies of the
rotational modes can be written by
$\kappa=\kappa_0+\kappa_2\bar\Omega^2$, the intercept $\kappa_0$ of
the fitting formula $y=\kappa_0+\kappa_2x$ can also be calculated by
a least-square method. We find that the coefficients $\eta_2$ and $\eta_2^\prime$ are in good
agreement with each other. 
It is important to note that the effects of the magnetic deformation
on the rotational modes
are quite small, which is similar to the case of low frequency
$g$-modes. We note that the frequency we obtain for $r$ mode is
consistent with that by Lander et al. (2010).
\begin{table*}
\begin{center}
\caption{Coefficient $\eta_2$ of $m=2$ rotational modes
  for isentropic polytropes with the indices $n=1$, 1.5, and 3  $^{*}$}
\label{symbols}
\begin{threeparttable}
\begin{tabular}{@{}crrr}
\hline
\hline
$l_0-|m|$ & $\kappa_0$ & $\eta_2$ & $\eta_2^\prime$ \\
\hline
&&$n=1$&\\
\hline
$1$ & 0.66666 & 8.324($-1$) & 8.326($-1$) \\
$2$ & -0.55660 & 9.105($-1$) & 9.171($-1$) \\
& 1.10002 & 2.398($-1$) & 2.401($-1$) \\
$3$ & -1.02590 & 3.304($-1$) & 3.316($-1$) \\
& 0.51734 & 1.784($+0$) & 1.785($+0$) \\
& 1.35777 & 1.402($-1$) & 1.404($-1$) \\
$4$ & -1.27290 & 2.481($-1$) & 2.511($-1$) \\
& -0.27533 & 7.103($+0$) & 7.100($+0$) \\
& 0.86296 & 5.805($-1$) & 5.820($-1$) \\
& 1.51956 & 9.950($-2$) & 9.957($-2$) \\
\hline
 &&$n=1.5$&\\
\hline
$1$ & 0.66666 & 5.247($-1$) & 5.244($-1$) \\
$2$ & -0.69650 & 3.688($-1$) & 3.696($-1$) \\
& 1.06257 & 1.752($-1$) & 1.753($-1$) \\
$3$ & -1.12782 & 1.439($-1$) & 1.443($-1$) \\
& 0.53564 & 1.071($+0$) & 1.069($+0$) \\
& 1.31001 & 1.103($-1$) & 1.104($-1$) \\
$4$ & -1.34198 & 8.378($-2$) & 8.382($-2$) \\
& -0.36425 & 2.999($+0$) & 2.993($+0$) \\
& 0.85864 & 3.516($-1$) & 3.520($-1$) \\
& 1.47217 & 8.210($-2$) & 8.237($-2$) \\
\hline
&&$n=3$&\\
\hline
$1$ & 0.66667 & 1.329($-1$) & 1.328($-1$) \\
$2$ & -1.07669 & 2.505($-2$) & 2.520($-2$) \\
& 0.99492 & 3.902($-2$) & 3.892($-2$) \\
$3$ & -1.37189 & 1.849($-2$) & 1.872($-2$) \\
& 0.57976 & 2.620($-1$) & 2.622($-1$) \\
& 1.20940 & 2.680($-2$) & 2.665($-2$) \\
$4$ & -1.51785 & 6.058($-3$) & 6.101($-3$) \\
& -0.66228 & 2.063($-1$) & 2.064($-1$) \\
& 0.85853 & 6.254($-2$) & 6.253($-2$) \\
& 1.36256 & 2.973($-2$) & 3.023($-2$) \\
\hline
\end{tabular}
\medskip
\begin{tablenotes}
\item[*] We use the notation of $1.000\times 10^N \equiv 1.000(N)$. 
\end{tablenotes}
\end{threeparttable}
\end{center}
\end{table*}

In Figure 5, we show the eigenfunctions of the $m=2$ rotational modes 
of the $n=1$ polytrope for $B_0=10^{16}$ G, where we assume $\bar\Omega=0.05$ and the amplitude normalization is given by
$T_{l^\prime_1}(R)=1$.
We find that the expansion coefficients for the $m=2$ inertial mode of
$\kappa_0(0)=1.1$ are the same as
those shown in figure 1 of Yoshida \& Lee (2000a). 
In Figure 6, we plot
the magnetic field perturbations $S_{bl}$, $H_{bl}$, and $T_{bl'}$ of the
$m=2$ rotational modes for
$B_0=10^{16}$ G. In Figures 7 and 8, we plot the eigenfunctions and
magnetic field perturbations of the $m=2$ rotational modes of the isentropic
polytrope with the index $n=3$ for $B_0=10^6$ G.

\begin{figure}
\begin{center}
\resizebox{0.39\columnwidth}{!}{
\includegraphics{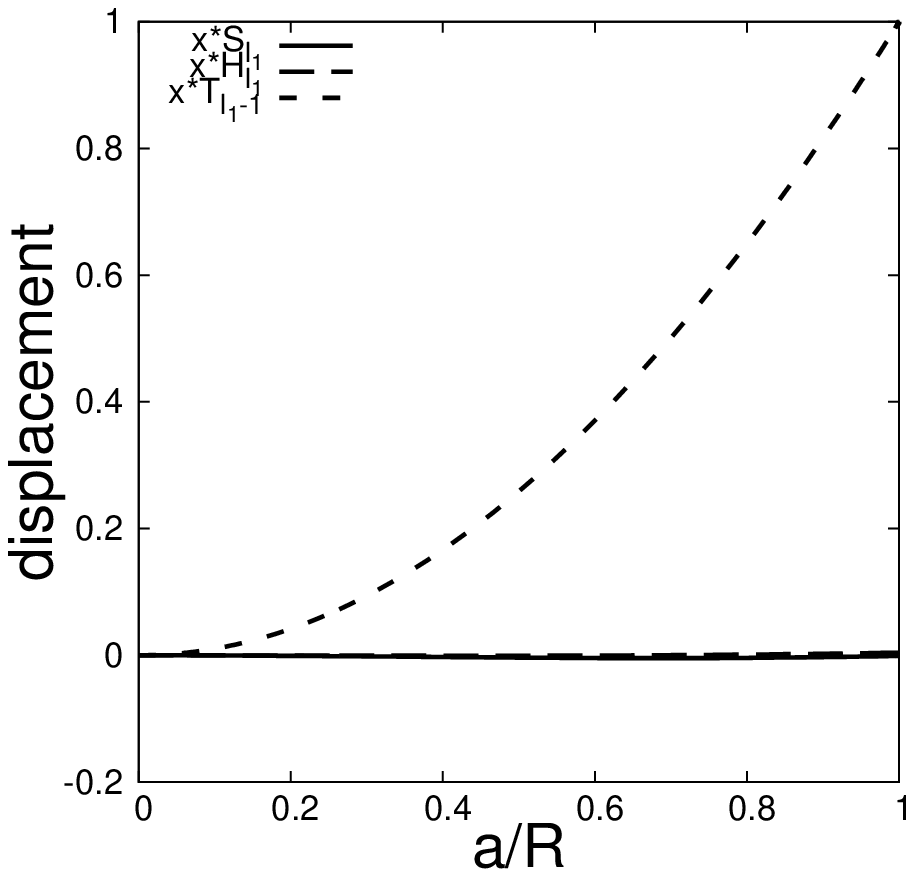}}
\hspace*{-1.75cm}
\resizebox{0.39\columnwidth}{!}{
\includegraphics{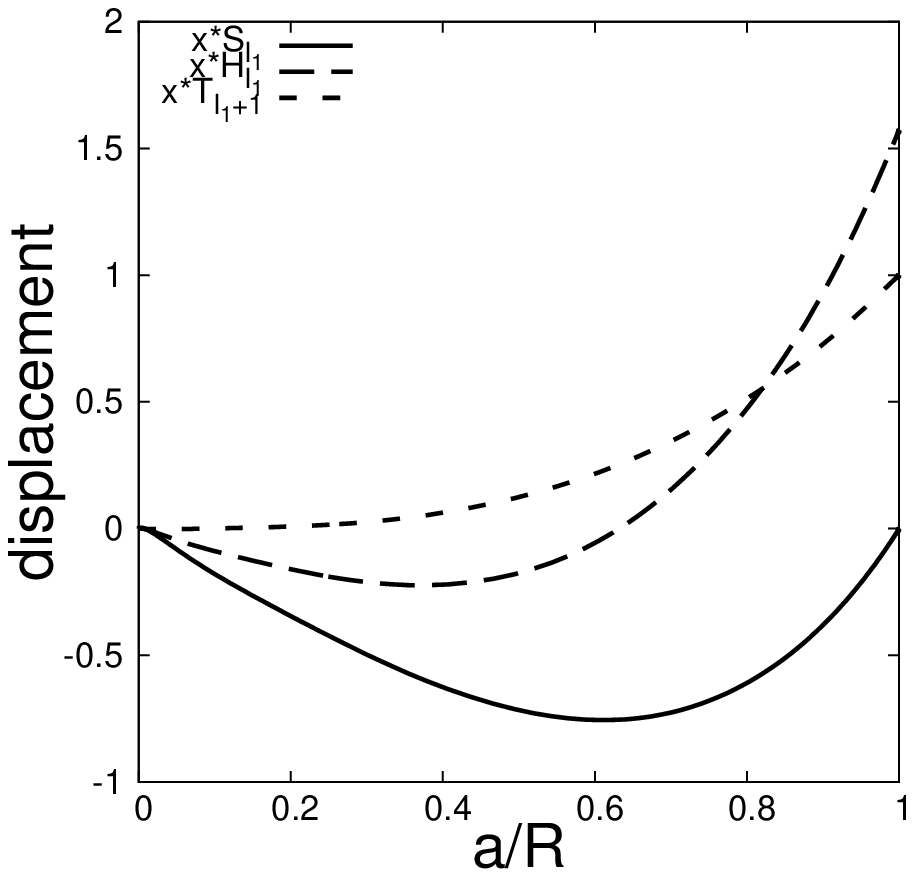}}
\hspace*{-1.75cm}
\resizebox{0.39\columnwidth}{!}{
\includegraphics{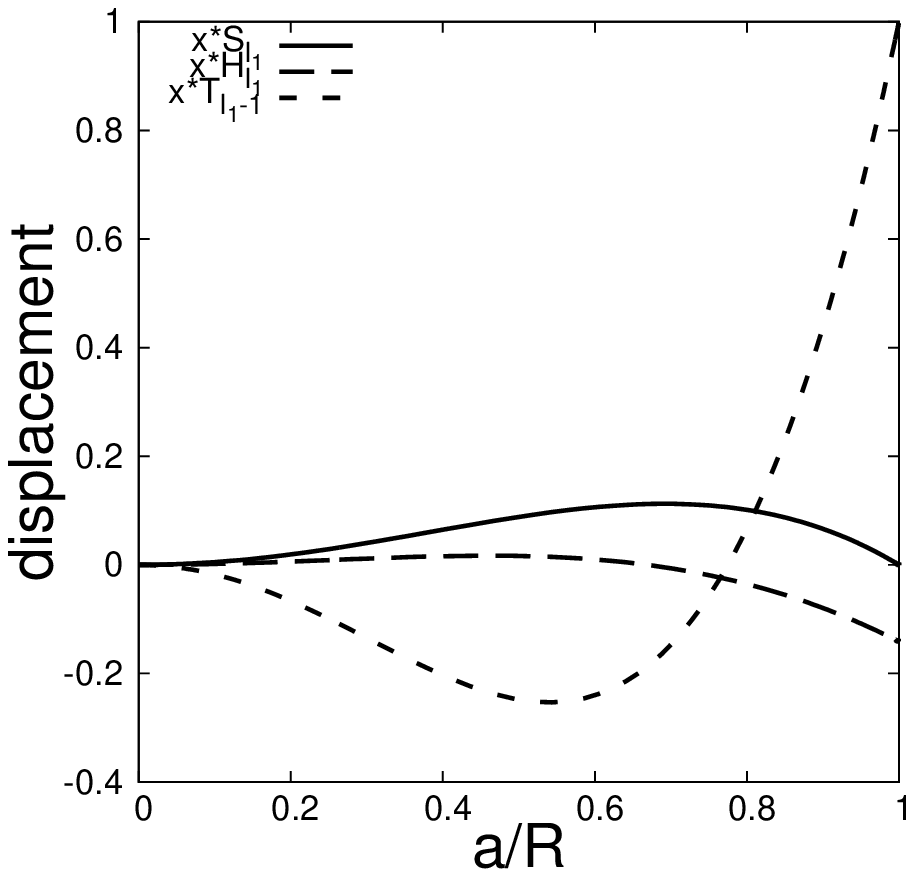}}
\end{center}
\caption{Eigenfunctions of $m=2$ rotational modes for the isentropic $n=1$ polytrope for $B_0=10^{16}$ G: 
$r$ mode of $\kappa_0=0.6667$ (left), and
  inertial modes of $\kappa_0=1.1000$ (center) and $\kappa_0=0.5173$ (right). 
  The solid lines, the long dashed lines, and the short dashed lines 
  are for the functions $xS_{l_1}$,
  $xH_{l_1}$, and $xT_{l_1^\prime}$ with $x=a/R$, respectively, and the amplitude normalization is given by ${\rm max}(xT_{l^\prime_1})=1$.}
\end{figure}

\begin{figure}
\begin{center}
\resizebox{0.39\columnwidth}{!}{
\includegraphics{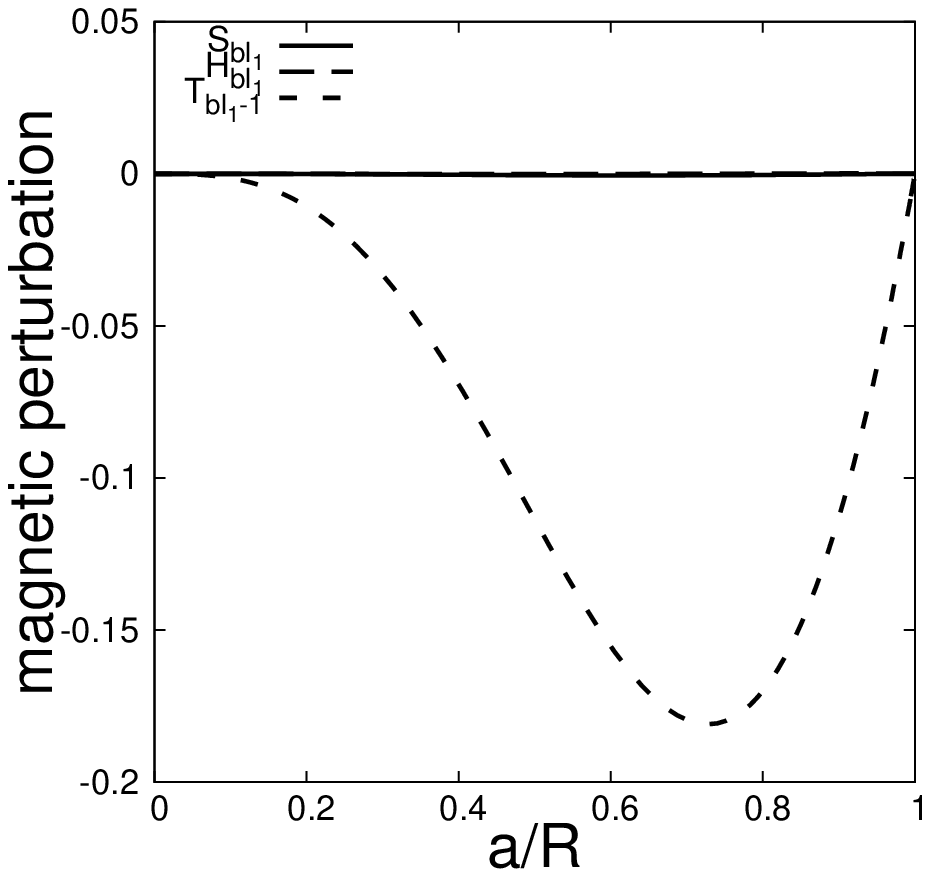}}
\hspace*{-1.75cm}
\resizebox{0.39\columnwidth}{!}{
\includegraphics{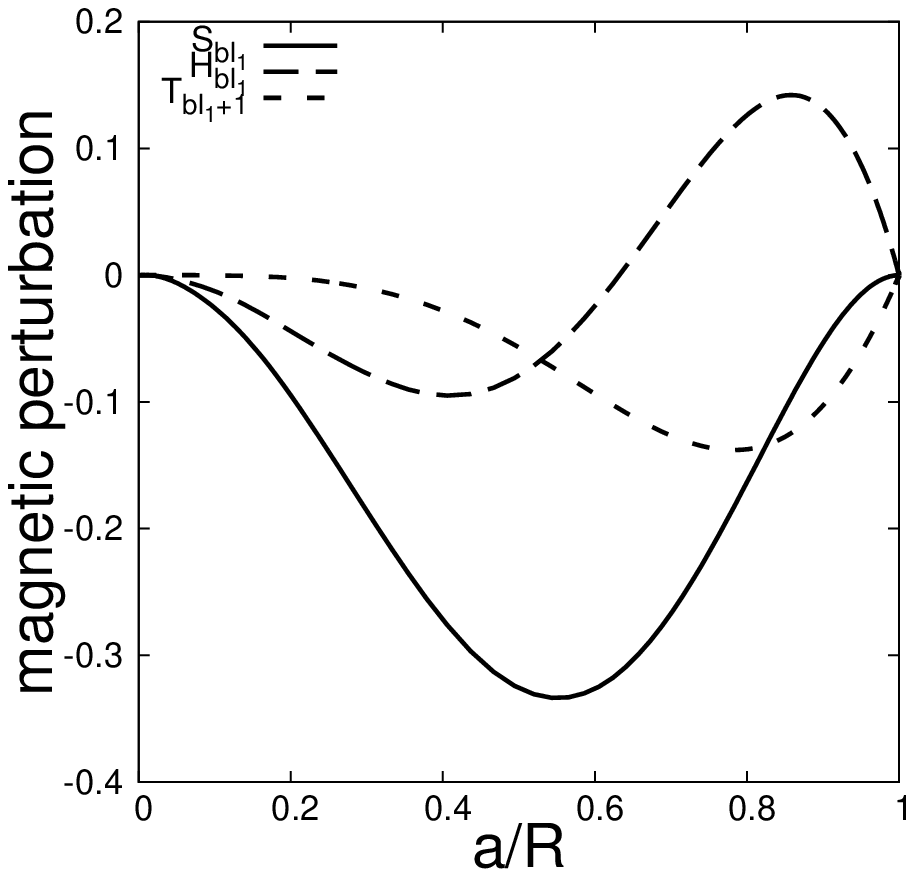}}
\hspace*{-1.75cm}
\resizebox{0.39\columnwidth}{!}{
\includegraphics{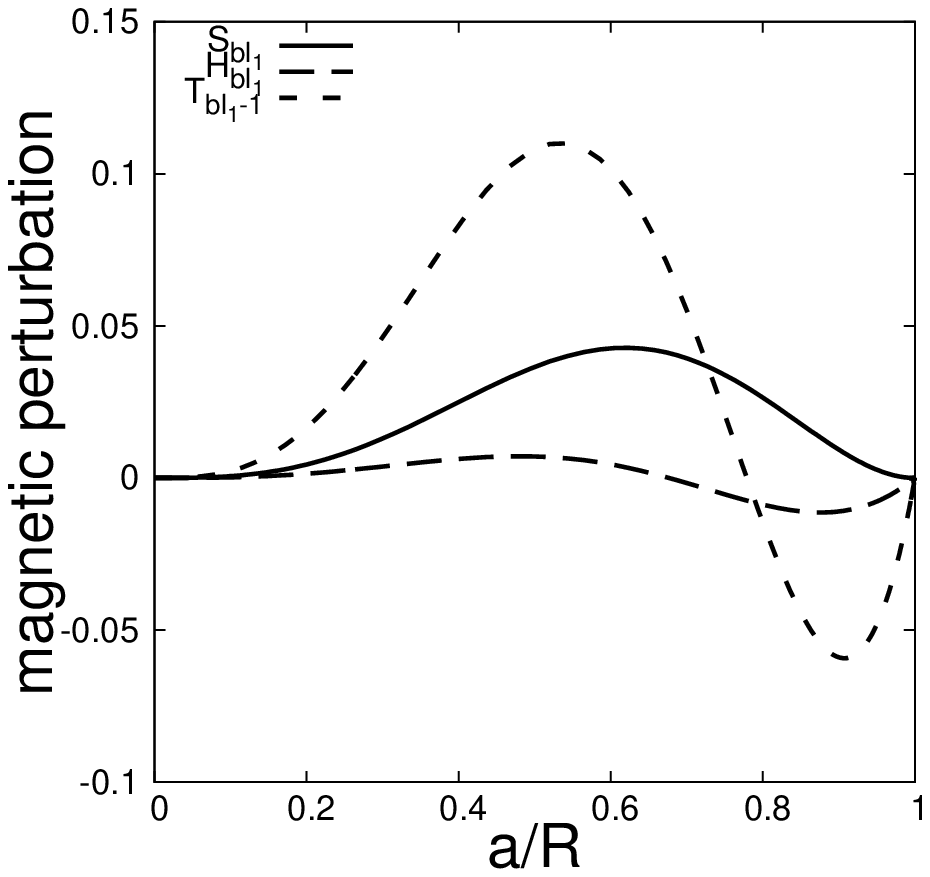}}
\end{center}
\caption{Same as Figure 5 but for the eigenfunctions
$S_{bl_1}\equiv k\rho a h_{l}^S/B_0$ (solid lines), $H_{bl_1}\equiv k\rho ah_{l}^H/B_0$ (long dashed lines), and 
$T_{bl_1^\prime}\equiv k\rho a  h_{l'}^T/B_0$ (short dashed lines).
}
\end{figure}

\begin{figure}
\begin{center}
\resizebox{0.39\columnwidth}{!}{
\includegraphics{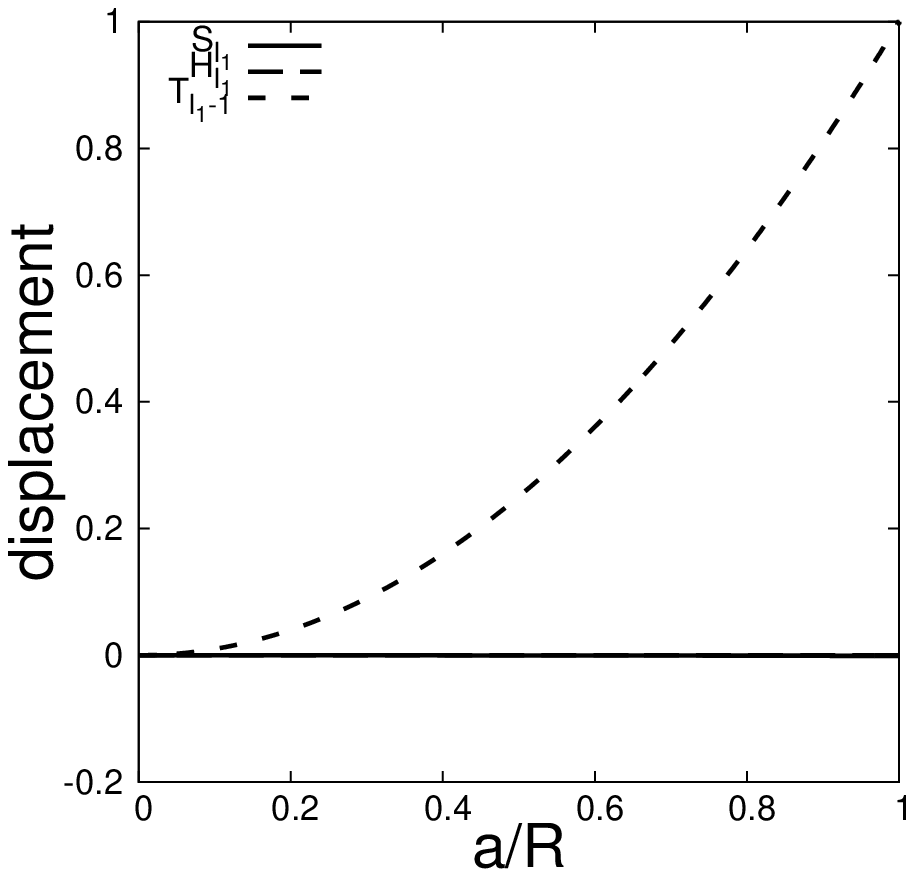}}
\hspace*{-1.75cm}
\resizebox{0.39\columnwidth}{!}{
\includegraphics{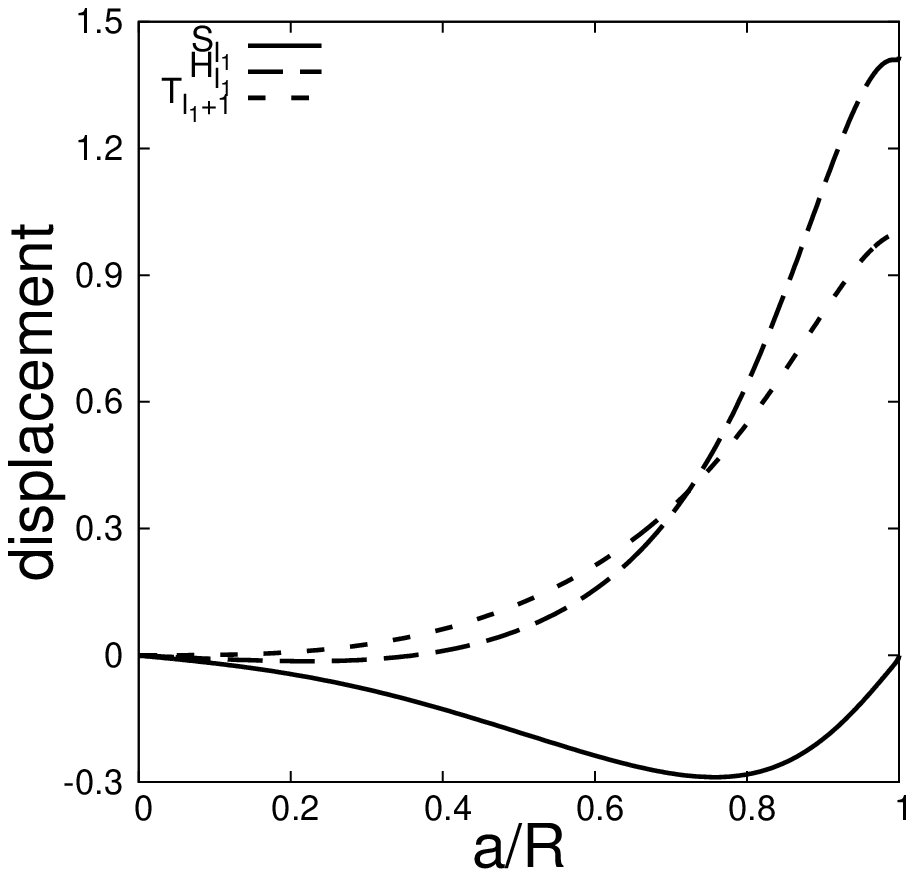}}
\hspace*{-1.75cm}
\resizebox{0.39\columnwidth}{!}{
\includegraphics{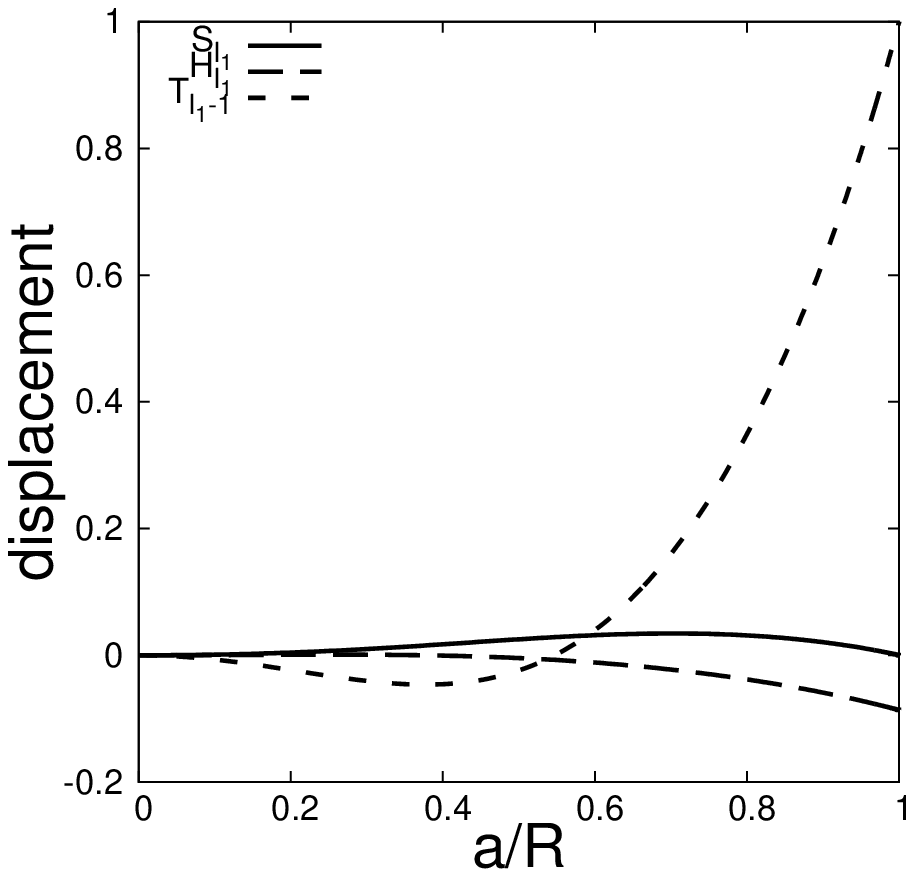}}
\end{center}
\caption{Eigenfunctions of $m=2$ rotational modes for the isentropic $n=3$ polytrope for $B_0=10^{6}$ G: 
r mode of $\kappa_0=0.6667$ (left), and
  inertial modes of $\kappa_0=0.9949$ (center) and $\kappa_0=0.5798$ (right). 
  The solid lines, the long dashed lines, and the short dashed lines 
  are for the functions $xS_{l_1}$,
  $xH_{l_1}$, and $xT_{l_1^\prime}$ with $x=a/R$, respectively, and the amplitude normalization is given by ${\rm max}(xT_{l^\prime_1})=1$.}
\end{figure}

\begin{figure}
\begin{center}
\resizebox{0.39\columnwidth}{!}{
\includegraphics{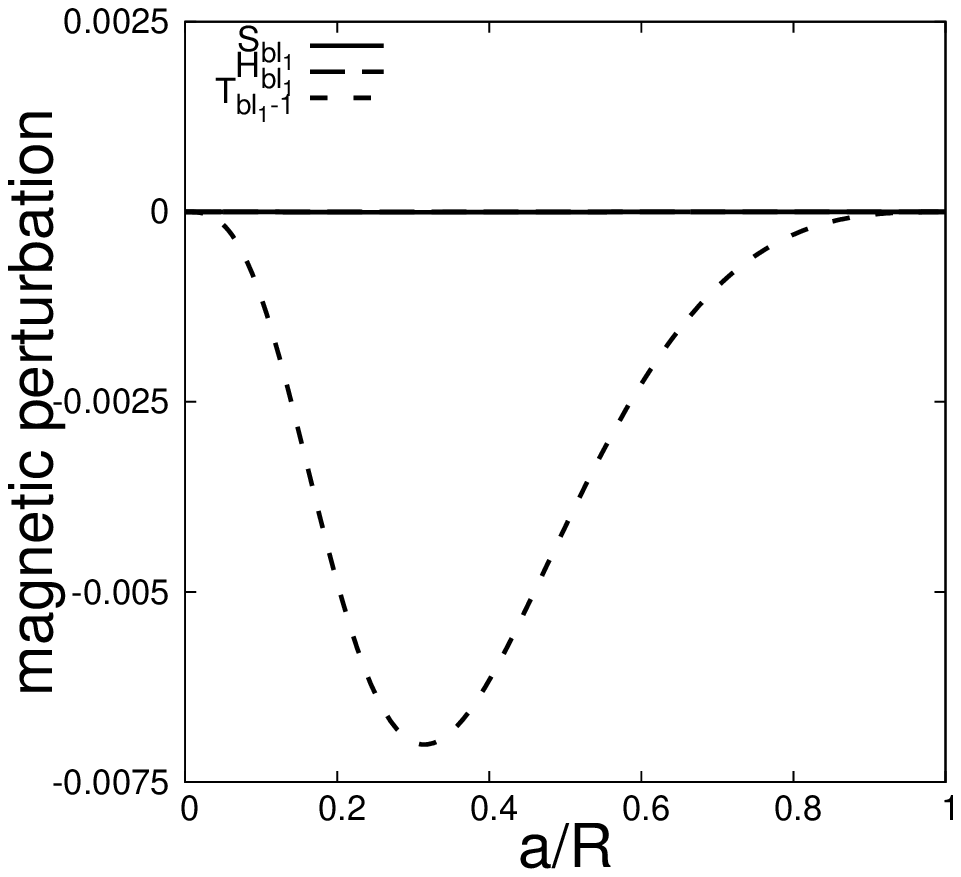}}
\hspace*{-1.75cm}
\resizebox{0.39\columnwidth}{!}{
\includegraphics{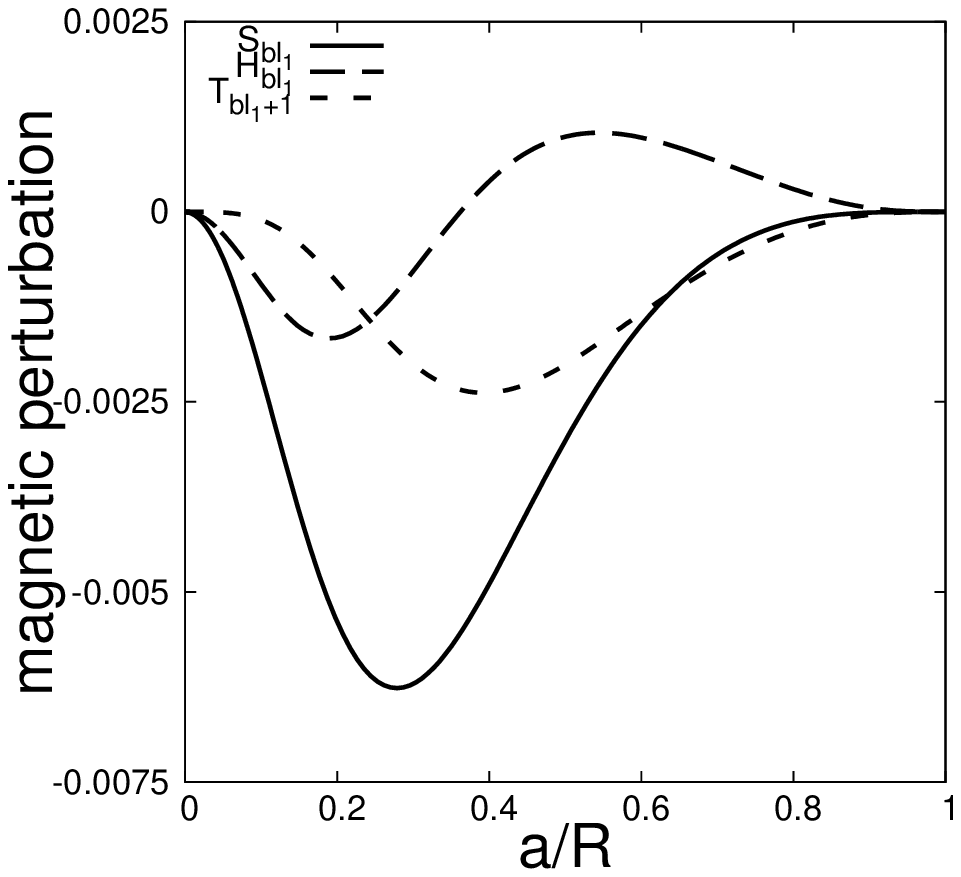}}
\hspace*{-1.75cm}
\resizebox{0.39\columnwidth}{!}{
\includegraphics{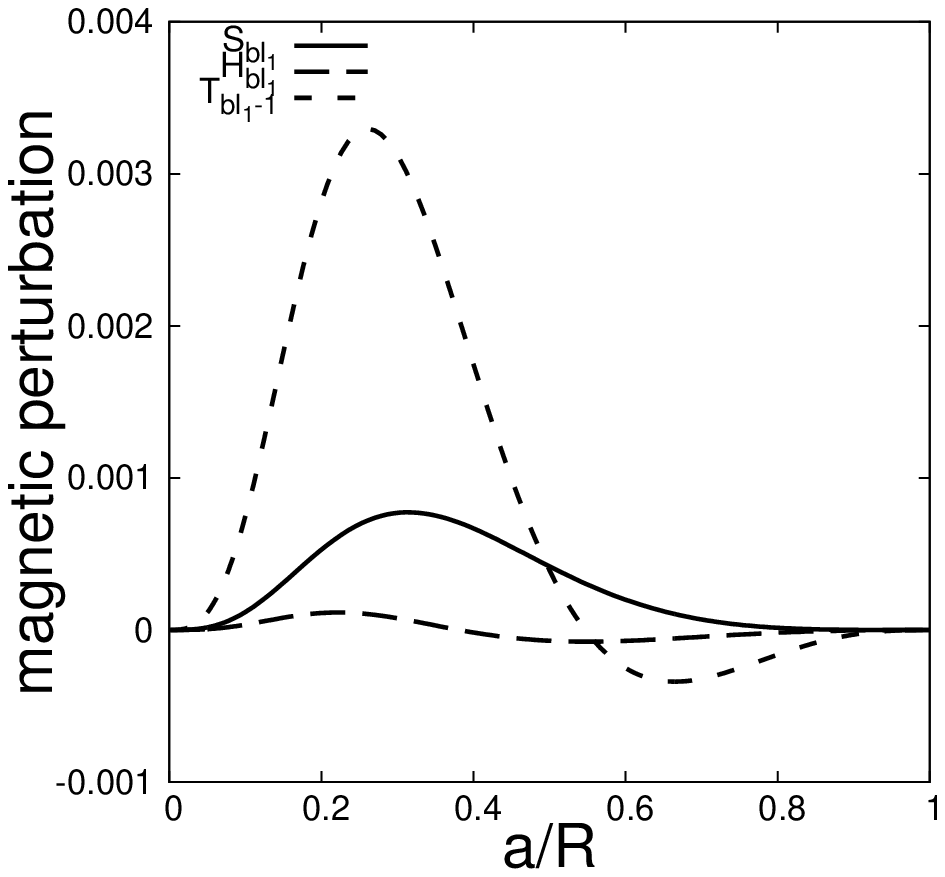}}
\end{center}
\caption{Same as Figure 7 but for the eigenfunctions
$S_{bl_1}\equiv k\rho a h_{l}^S/B_0$ (solid lines), $H_{bl_1}\equiv k\rho ah_{l}^H/B_0$ (long dashed lines), and 
$T_{bl_1^\prime}\equiv k\rho a  h_{l'}^T/B_0$ (short dashed lines).
}
\end{figure}

\subsection{magnetic modes}

We looked for very low frequency modes to find magnetic modes having real frequencies
for non-rotating stars, but we found none.
We obtained only solutions having pure imaginary $\sigma$ for a given value of $j_{\rm max}$, but 
we found that these solutions are dependent on $j_{\rm max}$ and cannot be regarded as 
correct solutions we look for. 
It is to be noted that we could not obtain very low frequency $g$-modes either
in the frequency range where magnetic modes having real frequencies might coexist.
The functions $\pmb{\xi}$ of the ``modes'' in that frequency range have discontinuities
as a function of $a$, which suggests that the ``modes'' are in a
continuum band of the frequency spectrum (see, section 7.4 of Goedbloed \& Poedts 2004).
This discontinuity of the functions $\pmb{\xi}$, which occurs in a certain
low frequency range, may be caused by
the relation $A\pmb{\Psi}=B\pmb{\Phi}$ coming from equations (A4) and (A5) and used to eliminate the variables $\pmb{\Psi}$
in equations (A2) and (A3), 
where $\pmb{\Psi}=(\pmb{H}$,$\pmb{T})^T$ and $\pmb{\Phi}=(\pmb{y}_1,\pmb{y}_2)^T$, and $A$ and $B$ are matrices,
that is, there appears a point at which the determinant of the matrix $A$ vanishes.
Goedbloed \& Poedts (2004) discussed the simplest case of a
second order ordinary differential equation that possesses a frequency band in which
the differential equation becomes singular at a point, and the situation we have in our calculations for
low frequency range is essentially the same as the second order differential equation.

This situation is largely different from that met by Lander et al. (2010), who
found polar and axial Alfv\'en modes for
magnetized rotating stars with a purely toroidal background magnetic
field. 
They suggested that pure Alfv\'en modes of a non-rotating star or
purely inertial modes of an unmagnetized rotating star may be replaced by
hybrid magneto-inertial modes for magnetized and rotating stars 
(see also Mathis \& Brye 2011, 2012), and that
in the limit of ${\cal M}/{\cal T}\to 0$ or $\bar{\omega}_A^2/\bar{\Omega}^2\to 0$ the hybrid modes reduce to purely inertial
modes, where $\cal M$ and $\cal T$ are the magnetic and kinetic energies of the equilibrium model,
respectively. 
The method of calculation Lander et al (2010) use is a MHD simulation that follows
the time development of linear oscillations around the equilibrium model and is different from the method we use in this paper. 
At this moment we do not understand why we find no magnetic modes for a purely
toroidal background magnetic field.

\section{Conclusion}
In this paper, we have calculated non-axisymmetric oscillations of
rotating stars magnetized with purely toroidal magnetic
fields. Here we have used polytropic models of the indices
$n=1$, $1.5$, and $3$, and included the effects of the deformation caused by the toroidal
magnetic fields. 
We have obtained discrete normal non-radial
oscillation modes such as $g$, $f$, and $p$ modes for non-rotating
case. 
The frequency change due to the magnetic field for the modes scale
with the square of typical stellar Alfv\'en frequency
$\omega_A\equiv\sqrt{B_0^2/(4\pi\rho_0R^2)}$, and the proportional coefficients
for the frequency changes are estimated by two different methods.
For rotating stars, we have obtained rotational modes such as $r$ and
inertial modes. 
The frequency changes for the rotational modes 
also scale with the square
of typical stellar Alfv\'en frequency $\omega_A$. 
From Tables 1 to 4, we find that  
high frequency modes such
as $f$ and $p$ modes are susceptible to the stellar deformation, while the
lower frequency modes such as $g$, $r$, and inertial modes are almost insensitive to the deformation.
It may be important to note that in the present analysis we could not find $j_{\rm max}$-independent
magnetic modes, the existence of which are  
suggested by Lander et al (2010).
The reason for the difference between the two calculations is not yet well understood.

The present analysis is a part of our study of the oscillations of
magnetized stars. Even for a purely poloidal magnetic field, it is
difficult to determine frequency spectra of non-axisymmetric modes and
those of axisymmetric spheroidal modes. We note that the stability of
a magnetic field configuration is another difficult problem. It is
well known that a purely poloidal and purely toroidal magnetic fields
are unstable and the energy of the field is dissipated quickly, that
is, for several ten milliseconds (e.g., Goosens 1979; Kiuchi, Yoshida, \& Shibata 2011; 
Laskey et al. 2011; Ciolfi \& Rezzolla 2012), although stellar rotation may weaken the instability
of a purely poloidal or purely toroidal magnetic fields (e.g., Lander
\& Jones 2011a, 2011b). It is thus anticipated that a mixed poloidal and
toroidal magnetic field configuration such as twisted-torus magnetic
field (e.g., Braithwaite \& Spruit 2004; Yoshida \& Eriguchi 2006; Yoshida, Yoshida \& Eriguchi
2006; Ciolfi et al. 2009) can be stable, that is, such a magnetic
field configuration can last stably for a long time. If this is the
case, it will be important to investigate the oscillation modes of
stars threded by both toroidal and poloidal magnetic fields. As
suggested by Colaiuda \& Kokkotas (2012), however, the presence of a
toroidal field component can significantly change the properties of
the oscillation modes of magnetized neutron stars. In the presence of
both poloidal and toroidal field components, toroidal and spheroidal
modes are coupled, which inevitably breaks equatorial symmetry and antisymmetry of perturbations.

As an important physical property inherent to cold neutron stars, we
need to consider the effects of superfluidity and superconductivity of
neutrons and protons on the oscillation modes (e.g., Andersson et
al. 2009; Galmpedakis et
al. 2011). It is believed that
neutrons become a superfluid both in the inner crust and in the fluid
core while protons can be superconducting in the core. For example, if
the fluid core is a type I superconductor, magnetic fields will be
expelled from the core region, because of the Meissner effect, and hence
confined to the solid crust (e.g., Colaiuda et al. 2008; Sotani et
al. 2008). In this case, we only have to consider oscillations of a
magnetized crust so long as toroidal modes are concerned, and we have
toroidal crust modes modified by a magnetic field, while spheroidal
oscillations can be propagative both in the magnetic crust and in the
non-magnetic fluid core. However, a recent analysis of the spectrum of
timing noise for SGR 1806-20 and SGR 1900+14 has suggested that the
core region is a type II superconductor (Arras, Cumming \& Thompson
2004). If this is the case, the fluid core can be threaded by a
magnetic field and hence the frequency spectra of oscillation modes
will be affected by the superconductivity in the core (e.g., Colaiuda
et al. 2008; Sotani et al. 2008).

\section*{Acknowledgements}
This work was supported in part by a Grant-in-Aid for Scientific Research from JSPS (No. 24540245).

\appendix

\section{Pulsation equations for the slowly rotating star with purely toroidal magnetic fields
  }
 
To describe the master equations concisely, it is useful to introduce the following column vectors composed of the expansion coefficients 
for the perturbation quantities: 
the vectors $\pmb{S}$, $\pmb{H}$, $\pmb{h}^S$, $\pmb{h}^S$, $\pmb{T}$, $\pmb{h}^T$, and $\pmb{y}_2$, defined by 
\begin{eqnarray}
(\pmb{S})_j=S_{l_j},\quad (\pmb{H})_j=H_{l_j},\quad
(\pmb{T})_j=T_{l'_j},\quad (\pmb{h}^S)_j=h^S_{l_j},\quad
(\pmb{h}^H)_j=h^H_{l_j},\quad (\pmb{h}^T)_j=h^T_{l'_j},\quad
(\pmb{y}_2)_j=\frac{p'_{l_j}}{\rho ag},
\label{eq:depvar}
\end{eqnarray}
where $(\pmb{X})_j$ denotes the $j$-th component of the vector $\pmb{X}$ and $g=GM(a)/a^2$ is the gravitational acceleration. 
The perturbed continuity equation (\ref{eq:continuityeq}), and the $a$--,  $\theta$-- and $\phi$--components of the perturbed Euler equation (\ref{eq:momentumeq}), 
respectively, reduce to 
\begin{eqnarray}
a\dfn{\pmb{S}}{a}{}=\left\{\left[V_G-3-a\dfn{\vartheta(\alpha)}{a}{}\right]\pmbmt{I}-a\dfn{\vartheta(\beta)}{a}{}\pmbmt{{\cal{A}}}_0\right\}\pmb{S}-V_G\pmb{y}_2+\left[\pmbmt{\Lambda}_0+3\vartheta(\beta)\pmbmt{{\cal{B}}}_0\right]\pmb{H}+3m\vartheta(\beta)\pmbmt{Q}_0\pmb{T} \,, 
\label{eq:conteq}
\end{eqnarray}
\begin{eqnarray}
a\dfn{\pmb{y}_2}{a}{}=\left[c_1\bar{\sigma}^2\left\{\left[1+2\eta(\alpha)\right]\pmbmt{I}+2\eta(\beta)\pmbmt{{\cal{A}}}_0\right\}
+aA\pmbmt{I}\right]\pmb{S}-{a A\over 3} \left(2+\dfn{\ln\rho}{\ln a}{}\right) \hat{\rho} c _1 {\bar{\omega}_A}^2 \left(\pmbmt{I}-\pmbmt{{\cal{A}}}_0\right) \pmb{S} 
\nonumber \\
+(1-aA-U)\pmb{y}_2+{V_G\over3}\left(2+\dfn{\ln\rho}{\ln a}{}\right) \hat{\rho} c _1 {\bar{\omega}_A}^2 \left(\pmbmt{I}-\pmbmt{{\cal{A}}}_0\right)\pmb{y}_2
\nonumber \\
-\{2mc_1\bar{\sigma}\bar{\Omega}[1+\alpha+\eta(\alpha)]\pmbmt{I}
+2mc_1\bar{\sigma}\bar{\Omega}[\beta+\eta(\beta)]\pmbmt{{\cal{A}}}_0+3c_1\bar{\sigma}^2\beta\pmbmt{{\cal{B}}}_0\}\pmb{H}
\nonumber \\
-\{2c_1\bar{\sigma}\bar{\Omega}[1+\alpha+\eta(\alpha)]\pmbmt{C}_0
+2c_1\bar{\sigma}\bar{\Omega}[\beta+\eta(\beta)]\pmbmt{{\cal{A}}}_0\pmbmt{C}_0+3mc_1\bar{\sigma}^2\beta\pmbmt{Q}_0\}\pmb{T}
\nonumber \\
+\frac{1}{2}\hat{\rho}c_1\bar{\omega}_A^2\left\{m\left[a\dfn{\pmb{h}^H}{a}{}-\pmb{h}^S+2\left(2+\dfn{\ln\rho}{\ln
        a}{}\right)\pmb{h}^H\right]-\pmbmt{C}_0\left[a\dfn{\pmb{h}^T}{a}{}+2\left(2+\dfn{\ln\rho}{\ln a}{}\right)\pmb{h}^T\right]\right\} \,, 
        \label{eq:moment_a}
\end{eqnarray}
\begin{eqnarray}
-\left\{m\nu\left[1+\alpha+\eta(\alpha)\right]\pmbmt{I}+m\nu\left[\beta+\eta(\beta)\right]\pmbmt{{\cal{A}}}_0-3\beta\left(2\pmbmt{{\cal{A}}}_0+\pmbmt{{\cal{B}}}_0\right)\right\}\pmb{S} +aA\hat{\rho}{{\bar{\omega}_A}^2\over \bar{\sigma}^2}\left(2\pmbmt{{\cal{A}}}_0+\pmbmt{{\cal{B}}}_0\right)\pmb{S}
\nonumber \\
-\frac{1}{c_1\bar{\sigma}^2}\pmbmt{\Lambda}_0\pmb{y}_2-\hat{\rho}{{\bar{\omega}_A}^2\over \bar{\sigma}^2}V_G\left(2\pmbmt{{\cal{A}}}_0+\pmbmt{{\cal{B}}}_0\right)\pmb{y}_2
\nonumber \\
+\left[\left(1+2\alpha\right)\pmbmt{\Lambda}_0\pmbmt{L}_0+2\beta\left(\pmbmt{{\cal{A}}}_0\pmbmt{\Lambda}_0+3\pmbmt{{\cal{B}}}_0\right)-2m\nu\beta\left(\pmbmt{I}+6\pmbmt{{\cal{A}}}_0\right)\right]\pmb{H}
\nonumber \\
+\left[-\nu\left(1+2\alpha-2\beta\right)\pmbmt{\Lambda}_0\pmbmt{M}_1-4\nu\beta\left(\pmbmt{{\cal{A}}}_0\pmbmt{\Lambda}_0\pmbmt{M}_1+3\pmbmt{Q}_0\pmbmt{{\cal{B}}}_1\right)+6m\beta\pmbmt{Q}_0\right]\pmb{T}
\nonumber \\
+\frac{1}{2}m\hat{\rho}\frac{\bar{\omega}_A^2}{\bar{\sigma}^2}\left(2+\dfn{\ln\rho}{\ln a}{}\right)\pmb{h}^S+m\hat{\rho}\frac{\bar{\omega}_A^2}{\bar{\sigma}^2}\pmb{h}^H-\frac{1}{2}\hat{\rho}\frac{\bar{\omega}_A^2}{\bar{\sigma}^2}\pmbmt{\Lambda}_0\pmbmt{R}\pmb{h}^T=0 \,, 
\label{eq:moment_theta}
\end{eqnarray}
\begin{eqnarray}
\left\{\nu\left[1+\alpha+\eta(\alpha)\right]\pmbmt{\Lambda}_1\pmbmt{K}+\nu\left[\beta+\eta(\beta)\right]\left(\pmbmt{{\cal{A}}}_1\pmbmt{\Lambda}_1\pmbmt{K}+3\pmbmt{Q}_1\pmbmt{Q}_0\pmbmt{Q}_1-3\pmbmt{Q}_1\right)-3m\beta\pmbmt{Q}_1\right\}\pmb{S}-maA\hat{\rho}{{\bar{\omega}_A}^2\over \bar{\sigma}^2}\pmbmt{Q}_1\pmb{S}
+m\hat{\rho}{{\bar{\omega}_A}^2\over \bar{\sigma}^2}V_G\pmbmt{Q}_1\pmb{y}_2
\nonumber \\
+\left[-\nu(1+2\alpha-2\beta)\pmbmt{\Lambda}_1\pmbmt{M}_0-4\nu\beta\left(\pmbmt{{\cal{A}}}_1\pmbmt{\Lambda}_1\pmbmt{M}_0+3\pmbmt{Q}_1\pmbmt{{\cal{B}}}_0\right)+6m\beta\pmbmt{Q}_1\right]\pmb{H}
\nonumber \\
+\left[(1+2\alpha)\pmbmt{\Lambda}_1\pmbmt{L}_1+2\beta\left(\pmbmt{{\cal{A}}}_1\pmbmt{\Lambda}_1+3\pmbmt{{\cal{B}}}_1\right)-2m\nu\beta\left(\pmbmt{I}+6\pmbmt{{\cal{A}}}_1\right)\right]\pmb{T}
\nonumber \\
-\frac{1}{2}\hat{\rho}\frac{\bar{\omega}_A^2}{\bar{\sigma}^2}\left(2+\dfn{\ln\rho}{\ln a}{}\right)\pmbmt{\Lambda}_1\pmbmt{K}\pmb{h}^S+\hat{\rho}\frac{\bar{\omega}_A^2}{\bar{\sigma}^2}\pmbmt{\Lambda}_1\pmbmt{M}_0\pmb{h}^H+\frac{1}{2}m\hat{\rho}\frac{\bar{\omega}_A^2}{\bar{\sigma}^2}\left(\pmbmt{\Lambda}_1-2\pmbmt{I}\right)\pmb{h}^T=0 \,. 
\label{eq:moment_phi}
\end{eqnarray}
The $a$--, $\theta$-- and $\phi$--components of the perturbed induction equation (\ref{eq:inductioneq}), respectvely, lead 
\begin{eqnarray}
\pmb{h}^S=m\pmb{S},
\label{eq:induc_a}
\end{eqnarray}
\begin{eqnarray}
\pmb{h}^H=maA\pmbmt{\Lambda}_0^{-1}\pmb{S}-mV_G\pmbmt{\Lambda}_0^{-1}\pmb{y}_2+m\pmb{H},
\label{eq:induc_theta}
\end{eqnarray}
\begin{eqnarray}
\pmb{h}^T=aA\pmbmt{K}\pmb{S}-V_G\pmbmt{K}\pmb{y}_2-m\pmb{T}.
\label{eq:induc_phi}
\end{eqnarray}
Here,
\begin{eqnarray}
U=\dfn{\ln M(a)}{\ln a}{},\quad V_G=-\frac{1}{\Gamma_1}\dfn{\ln p}{\ln a}{},\quad
\vartheta(\alpha)=3\alpha+a\dfn{\alpha}{a}{},\quad \eta(\alpha)=\alpha+a\dfn{\alpha}{a}{} \,, 
\end{eqnarray}
and $\bar{\sigma}\equiv\sigma/(GM/R^3)^{1/2}$ is the frequency in the unit
of the Kepler frequency, and $\nu\equiv 2\Omega/\sigma$. The quantities $\pmbmt{Q}_0$, $\pmbmt{Q}_1$, $\pmbmt{C}_0$,
$\pmbmt{C}_1$, $\pmbmt{K}$, $\pmbmt{M}_0$, $\pmbmt{M}_1$, $\pmbmt{\Lambda}_0$,
$\pmbmt{\Lambda_1}$, $\pmbmt{R}$, $\pmbmt{L}_0$, $\pmbmt{L}_1$,
$\pmbmt{{\cal{A}}}_0$, $\pmbmt{{\cal{A}}}_1$, $\pmbmt{{\cal{B}}}_0$,
and $\pmbmt{{\cal{B}}}_1$ denote the matrices defined as follows: 

\noindent
For even modes, 
\begin{eqnarray}
(\pmbmt{Q}_0)_{jj}=J_{l_j+1}^m, \quad
(\pmbmt{Q}_0)_{j+1,j}=J_{l_j+2}^m,\quad
(\pmbmt{Q}_1)_{jj}=J_{l_j+1}^m,\quad
(\pmbmt{Q}_1)_{j,j+1}=J_{l_j+2}^m, \nonumber
\end{eqnarray}
\begin{eqnarray}
(\pmbmt{C}_0)_{jj}=-(l_j+2)J_{l_j+1}^m,\quad
(\pmbmt{C}_0)_{j+1,j}=(l_j+1)J_{l_j+2}^m, \quad
(\pmbmt{C}_1)_{jj}=l_jJ_{l_j+1}^m,\quad
(\pmbmt{C}_1)_{j,j+1}=-(l_j+3)J_{l_j+2}^m, \nonumber
\end{eqnarray}
\begin{eqnarray}
(\pmbmt{K})_{jj}=\frac{J_{l_j+1}^m}{l_j+1},\quad
(\pmbmt{K})_{j,j+1}=-\frac{J_{l_j+2}^m}{l_j+2}, \nonumber
\end{eqnarray}
\begin{eqnarray}
(\pmbmt{M}_0)_{jj}=\frac{l_j}{l_j+1}J_{l_j+1}^m,\quad
(\pmbmt{M}_0)_{j,j+1}=\frac{l_j+3}{l_j+2}J_{l_j+2}^m,\quad
(\pmbmt{M}_1)_{jj}=\frac{l_j+2}{l_j+1}J_{l_j+1}^m, \quad
(\pmbmt{M}_1)_{j+1,j}=\frac{l_j+1}{l_j+2}J_{l_j+2}^m, \nonumber
\end{eqnarray}
\begin{eqnarray}
(\pmbmt{\Lambda}_0)_{jj}=l_j(l_j+1),\quad
(\pmbmt{\Lambda}_1)_{jj}=(l_j+1)(l_j+2), \nonumber
\end{eqnarray}
\begin{eqnarray}
(\pmbmt{R})_{jj}=-\frac{(l_j+2)(l_j-1)}{l_j+1}J_{l_j+1}^m, \quad
(\pmbmt{R})_{j+1,j}=\frac{(l_j+1)(l_j+4)}{l_j+2}J_{l_j+2}^m, \nonumber
\end{eqnarray}
\begin{eqnarray}
\pmbmt{L}_0=\pmbmt{I}-m\nu\pmbmt{\Lambda}_0^{-1},\quad
\pmbmt{L}_1=\pmbmt{I}-m\nu\pmbmt{\Lambda}_1^{-1}, \quad
\pmbmt{{\cal{A}}}_0=\frac{1}{2}(3\pmbmt{Q}_0\pmbmt{Q}_1-\pmbmt{I}),\quad
\pmbmt{{\cal{A}}}_1=\frac{1}{2}(3\pmbmt{Q}_1\pmbmt{Q}_0-\pmbmt{I}),
\quad \pmbmt{{\cal{B}}}_0=\pmbmt{Q}_0\pmbmt{C}_1,\quad \pmbmt{{\cal{B}}}_1=\pmbmt{Q}_1\pmbmt{C}_0,
\end{eqnarray}
where $l_j=|m|+2j-2$ for $j=1,2,3,....,j_{\rm max}$, and 
\begin{eqnarray}
J_{l_j}^m=\left[\frac{(l_j+m)(l_j-m)}{(2l_j-1)(2l_j+1)}\right]^{1/2}.
\end{eqnarray}
For odd modes,
\begin{eqnarray}
(\pmbmt{Q}_0)_{jj}=J_{l_j+1}^m,\quad
(\pmbmt{Q}_0)_{j,j+1}=J_{l_j+2}^m, \quad
(\pmbmt{Q}_1)_{jj}=J_{l_j+1}^m,\quad
(\pmbmt{Q}_1)_{j+1,j}=J_{l_j+2}^m, \nonumber
\end{eqnarray}
\begin{eqnarray}
(\pmbmt{C}_0)_{jj}=l_jJ_{l_j+1}^m,\quad
(\pmbmt{C}_0)_{j,j+1}=-(l_j+3)J_{l_j+2}^m, \quad
(\pmbmt{C}_1)_{jj}=-(l_j+2)J_{l_j+1}^m,\quad
(\pmbmt{C}_1)_{j+1,j}=(l_j+1)J_{l_j+2}^m, \nonumber
\end{eqnarray}
\begin{eqnarray}
(\pmbmt{K})_{jj}=-\frac{J_{l_j+1}^m}{l_j+1},\quad
(\pmbmt{K})_{j+1,j}=\frac{J_{l_j+2}^m}{l_j+2}, \nonumber
\end{eqnarray}
\begin{eqnarray}
(\pmbmt{M}_0)_{jj}=\frac{l_j+2}{l_j+1}J_{l_j+1}^m, \quad
(\pmbmt{M}_0)_{j+1,j}=\frac{l_j+1}{l_j+2}J_{l_j+2}^m, \quad
(\pmbmt{M}_1)_{jj}=\frac{l_j}{l_j+1}J_{l_j+1}^m,\quad
(\pmbmt{M}_1)_{j,j+1}=\frac{l_j+3}{l_j+2}J_{l_j+2}^m, \nonumber
\end{eqnarray}
\begin{eqnarray}
(\pmbmt{\Lambda}_0)_{jj}=(l_j+1)(l_j+2),\quad
(\pmbmt{\Lambda}_1)_{jj}=l_j(l_j+1), \nonumber
\end{eqnarray}
\begin{eqnarray}
(\pmbmt{R})_{jj}=\frac{l_j(l_j+3)}{l_j+1}J_{l_j+1}^m,\quad
(\pmbmt{R})_{j,j+1}=-\frac{l_j(l_j+3)}{l_j+2}J_{l_j+2}^m, \nonumber
\end{eqnarray}
\begin{eqnarray}
\pmbmt{L}_0=\pmbmt{I}-m\nu\pmbmt{\Lambda}_0^{-1},\quad
\pmbmt{L}_1=\pmbmt{I}-m\nu\pmbmt{\Lambda}_1^{-1}, \quad
\pmbmt{{\cal{A}}}_0=\frac{1}{2}(3\pmbmt{Q}_0\pmbmt{Q}_1-\pmbmt{I}),\quad
\pmbmt{{\cal{A}}}_1=\frac{1}{2}(3\pmbmt{Q}_1\pmbmt{Q}_0-\pmbmt{I}),
\quad \pmbmt{{\cal{B}}}_0=\pmbmt{Q}_0\pmbmt{C}_1,\quad \pmbmt{{\cal{B}}}_1=\pmbmt{Q}_1\pmbmt{C}_0,
\end{eqnarray}
where $l_j=|m|+2j-1$ for $j=1,2,3,....,j_{\rm max}$.

Eliminating the variables $\pmb{h}^S$, $\pmb{h}^H$, and $\pmb{h}^T$ from equations
(\ref{eq:moment_a})-(\ref{eq:moment_phi})
by using equations (\ref{eq:induc_a})-(\ref{eq:induc_phi}), and eliminating $\pmb{H}$ and
$\pmb{T}$ 
by using equations (\ref{eq:moment_theta}) and
(\ref{eq:moment_phi}), we may reduce equations (\ref{eq:conteq}) and (\ref{eq:moment_a}) to a set of coupled first-order linear
ordinary differential equations for the functions $\pmb{y}_1=\pmb{S}$ and $\pmb{y}_2$, which is formally written as: 
\begin{eqnarray}
a\dfn{}{a}{}\left(\begin{array}{c}
\pmb{y}_1 \\
\pmb{y}_2
\end{array}\right)=\pmbmt{{\cal{F}}}\left(\begin{array}{c}
\pmb{y}_1 \\
\pmb{y}_2
\end{array}\right).
\end{eqnarray}
The surface boundary conditions are
\begin{eqnarray}
-\pmb{y}_1+\pmb{y}_2
+\hat{\rho}c_1\bar{\omega}_A^2\left[{1\over 3}\left(1+\dfn{\ln\rho}{\ln a}{}\right)\left(\pmbmt{I}-\pmbmt{{\cal{A}}}_0\right) \pmb{S} 
+{1\over2}\pmbmt{{\cal{B}}}_0\,\pmb{H} +{1\over2}m\pmbmt{Q}_0\pmb{T}-{1\over2}m\pmb{h}^H +{1\over2}\pmbmt{{\cal{C}}}_0\,\pmb{h}^T \right]
=0,
\label{eq:surfaceboundary}
\end{eqnarray}
which means $\displaystyle {1\over pV}\Delta\left(p+{1\over 8\pi}|B|^2 \right)=0$ at the stellar surface, where $\Delta Q$ denotes 
the Lagrangian change of the quantity $Q$. The boundary
conditions at the stellar center are the regularity conditions for the eigenfunctions $\pmb{y}_1$ and $\pmb{y}_2$.

\section{surface boundary condition for the function $\psi_2$}

Te determine the function $\psi_2$ satisfying the differential equation (\ref{eq:diffpsi2}), 
we need the surface boundary conditions.
Assuming the deviation of the surface $r=R_s(R,\theta)$ of the magnetized star from the surface $r=R$
of the non-magnetized star is small, we may write
\begin{eqnarray}
R_s(R,\theta)=R\left(1+\delta\zeta(\theta)\right).
\end{eqnarray}
Since we have $\rho(R_s,\theta)=0$ and $\rho_0(R)=0$ at the stellar surface, from equation (\ref{eq:rho_expand}) we can obtain 

\begin{eqnarray}
\delta\zeta=2R^2\omega_A^2\dfn{x}{\Psi_0}{}\left[\psi_0(1)+\psi_2(1)P_2(\cos\theta)\right].
\end{eqnarray}
Using equation (\ref{eq:quasipot}), the gravitational potential $\Phi(r,\theta)$ and
its derivative $\partial\Phi(r,\theta)/\partial x$ inside the star
are given by
\begin{eqnarray}
\Phi(r,\theta)=\Psi_0(r)+c_0-2R^2\omega_A^2\left[c_{1,0}+\psi_0(x)+\psi_2(x)P_2(\cos\theta)\right]-\frac{1}{3}\omega_A^2r^2\hat{\rho}\left[1-P_2(\cos\theta)\right],
\end{eqnarray}

\begin{eqnarray}
\pdn{\Phi(r,\theta)}{x}{}=\pdn{\Psi_0(r)}{x}{}-2R^2\omega_A^2\left[{d\psi_0\over dx}(x)+{d\psi_2\over dx}(x)P_2(\cos\theta)\right]
-\frac{1}{3}\omega_A^2\left(r^2{d\hat{\rho}\over dx}+2Rr\hat{\rho}\right)\left[1-P_2(\cos\theta)\right],
\end{eqnarray}
where we have set the constant $C$ in equation (\ref{eq:quasipot}) as $C=c_0-2R^2\omega_A^2c_{1,0}$. 
Since $\hat\rho(R_s,\theta)=0$ and $\Psi_0(R_s)\approx\Psi_0(R)+(d\Psi_0/dr)_{r=R}R\delta\zeta$
and $(\partial\Psi_0(r)/\partial r)_{r=R_s}=GM/R_s^2\approx (GM/R^2)(1-2\delta\zeta)$ at the deformed surface, the gravitational potential 
$\Phi(R_s,\theta)$ 
and its derivative $\partial\Phi(R_s,\theta)/\partial x$ reduce to
\begin{eqnarray}
\Phi(R_s,\theta)
=\Psi_0(R)+c_0-2R^2\omega_A^2c_{1,0},
\end{eqnarray}
\begin{eqnarray}
{\partial\Phi\over\partial x}(R_s,\theta)=\pdn{\Psi_0(R)}{x}{}-2R^2\omega_A^2\left[{d\psi_0\over dx}(1)+{d\psi_2\over dx}(1)P_2(\cos\theta)\right]
-\frac{1}{3}\omega_A^2R^2{d\hat{\rho}\over
  dx}\left[1-P_2(\cos\theta)\right]-4R^2\omega_A^2\left[\psi_0(1)+\psi_2(1)P_2(\cos\theta)\right], \nonumber
\end{eqnarray}
\begin{equation}
\end{equation}
where we have used $d\Psi_0/dx=GM/R^2$ at the surface.
On the other hand, the gravitational potential outside the star can be
written as
\begin{eqnarray}
\Phi=-\frac{\kappa_0}{x}-2R^2\omega_A^2\left[\frac{\kappa_{1,0}}{x}+\frac{\kappa_{1,2}}{x^3}P_2(\cos\theta)\right],
\end{eqnarray}
\begin{eqnarray}
{\partial\Phi\over\partial x}=\frac{\kappa_0}{x^2}+2R^2\omega_A^2\left[\frac{\kappa_{1,0}}{x^2}+3\frac{\kappa_{1,2}}{x^4}P_2(\cos\theta)\right],
\end{eqnarray}
and at the stellar surface $x=x_s\equiv 1+\delta\zeta$ we have
\begin{eqnarray}
\Phi=-\kappa_0-2R^2\omega_A^2\left[\kappa_{1,0}+\kappa_{1,2}P_2(\cos\theta)\right]+\kappa_0\delta\zeta,
\end{eqnarray}
\begin{eqnarray}
{\partial\Phi\over\partial x}=\kappa_0+2R^2\omega_A^2\left[\kappa_{1,0}+3\kappa_{1,2}P_2(\cos\theta)\right]-2\kappa_0\delta\zeta, 
\end{eqnarray}
where $\kappa_0$, $\kappa_{1,0}$, and
$\kappa_{1,2}$ are arbitrary constants. 
If we require $\Phi$ and
$\partial\Phi/\partial x$ inside and outside the star are continuous at the stellar surface, by comparing
the zeroth-order terms, we find
\begin{eqnarray}
\Psi_0(R)+c_0=-\kappa_0, \quad \kappa_0=\pdn{\Psi_0(R)}{x}{},
\end{eqnarray}
and hence we obtain
\begin{eqnarray}
\Psi_0(R)=-\kappa_0=-\pdn{\Psi_0(R)}{x}{}, \quad c_0=0.
\end{eqnarray}
By comparing the perturbed terms we can obtain following
relations:
\begin{eqnarray}
-c_{1,0}=-\kappa_{1,0}+\psi_0(1),\quad \kappa_{1,2}=\psi_2(1),
\end{eqnarray}
\begin{eqnarray}
\kappa_{1,0}=-{d\psi_0\over dx}(1)-\frac{1}{6}{d\hat{\rho}\over dx}(1),\quad 3\kappa_{1,2}=-{d\psi_2\over dx}(1)+\frac{1}{6}{d\hat{\rho}\over dx}(1).
\end{eqnarray}
From the relations, we note that 
the unknown constants $c_{1,0}$ and $\kappa_{1,0}$ for the function $\psi_0$ are determined uniquely by integrating equation (\ref{eq:diffpsi0}) to the surface:
\begin{eqnarray}
\kappa_{1,0}=-{d\psi_0\over dx}(1)-\frac{1}{6}{d\hat{\rho}\over dx}(1),\quad c_{1,0}=-{d\psi_0\over dx}(1)-\psi_0(1)-\frac{1}{6}{d\hat{\rho}\over dx}(1).
\end{eqnarray}
On the other hand, for the function $\psi_2$ we obtain by eliminating the constant $\kappa_{1,2}$
\begin{eqnarray}
3\psi_2(1)+{d\psi_2\over dx}(1)=\frac{1}{6}{d\hat{\rho}\over dx}(1),
\end{eqnarray}
which gives the outer boundary condition for $\psi_2$ at the surface.

\section{Frequency changes due to the magnetic fields}

Using the continuity equation (\ref{eq:continuityeq}) and the adiabatic relation (\ref{eq:rhoprime}), we may rewrite
the Euler equation (\ref{eq:momentumeq}) as  
\begin{eqnarray}
-\sigma^2[\left(1+2\epsilon\right)\pmb{\xi}+a\xi^a\nabla_0\epsilon+a(\pmb{\xi}\cdot\nabla_0\epsilon)\pmb{e}_a]=-\nabla_0\chi+\pmb{e}_a\frac{\Gamma_1p}{\rho}A\left[\nabla_0\cdot\pmb{\xi}+\pmb{\xi}\cdot\nabla_0\left(3\epsilon+a\pdn{\epsilon}{a}{}\right)\right]+i\sigma\pmb{D}
\nonumber \\
+\frac{\left(\pmb{\xi}\cdot\nabla_0\ln{\rho}+\nabla_0\cdot\pmb{\xi}\right)}{4\pi\rho}\left(\nabla_0\times\pmb{B}\right)\times\pmb{B}
+\frac{1}{4\pi\rho}\left[\left(\nabla_0\times\pmb{B}\right)\times\pmb{B}^\prime+\left(\nabla_0\times\pmb{B}^\prime\right)\times\pmb{B}\right],
\end{eqnarray}
where $\chi\equiv p^\prime/\rho$. We write the eigenfunctions and eigenfrequency as follows (for a similar treatment, see, e.g., Saio 1981): 
\begin{eqnarray}
\pmb{\xi}=\pmb{\xi}_0+\pmb{\xi}_2,
\end{eqnarray}
\begin{eqnarray}
\chi=\chi_0+\chi_2,
\end{eqnarray}
\begin{eqnarray}
\sigma=\sigma_0+\sigma_2,
\end{eqnarray}
where quantities with subscripts 0 and 2 denote quantities of order $\omega_A^0$ and $\omega_A^2$,
respectively. The Coriolis term, $\pmb{D}$, is then written by  
\begin{eqnarray}
\pmb{D}=\pmb{D}^{(0)}[\pmb{\xi}_0]+\pmb{D}^{(0)}[\pmb{\xi}_2]+\pmb{D}^{(2)}[\pmb{\xi}_0] ,
\end{eqnarray}
where 
\begin{eqnarray}
&&D^{(0)}_a[\pmb{\xi}]=2\Omega \sin\theta\xi^\phi,
\quad
D^{(0)}_\theta[\pmb{\xi}]=2\Omega \cos\theta\xi^\phi, 
\quad 
D^{(0)}_\phi[\pmb{\xi}]=-2\Omega\left( \sin\theta\xi^a+\cos\theta\xi^\theta\right),  \nonumber
\\
&&D^{(2)}_a[\pmb{\xi}]=2\Omega\left(2\epsilon+a\pdn{\epsilon}{a}{}\right)\sin\theta\xi^\phi,
\quad
D^{(2)}_\theta[\pmb{\xi}]=2\Omega\left(2\epsilon+\frac{\sin\theta}{\cos\theta}\pdn{\epsilon}{\theta}{}\right)\cos\theta\xi^\phi, \nonumber \\
&&D^{(2)}_\phi[\pmb{\xi}]=-2\Omega\left[\left(2\epsilon+a\pdn{\epsilon}{a}{}\right)\sin\theta\xi^a
+\left(2\epsilon+\frac{\sin\theta}{\cos\theta}\pdn{\epsilon}{\theta}{}\right)\cos\theta\xi^\theta\right]\,.  \nonumber
\end{eqnarray}
Introducing equations (C2)-(C5) into equation (C1) and
grouping quantities of the same order in $\omega_A$, we obtain
\begin{eqnarray}
-\sigma_0^2\pmb{\xi}_0=-\nabla_0\chi_0+\pmb{e}_a\frac{\Gamma_1p}{\rho}A\nabla_0\cdot\pmb{\xi}_0+i\sigma_0\pmb{D}^{(0)}[\pmb{\xi}_0] \,.
\end{eqnarray}
for order $\omega_A^0$, and
\begin{eqnarray}
-\sigma_0^2\pmb{\xi}_2-2\sigma_0^2\epsilon\pmb{\xi}_0-\sigma_0^2a\xi^a_0\nabla_0\epsilon-\sigma_0^2(\pmb{\xi}_0\cdot\nabla_0\epsilon)\pmb{e}_a
-2\sigma_0\sigma_2\pmb{\xi}_0
\nonumber \\
=-\nabla_0\chi_2+\pmb{e}_a\frac{\Gamma_1p}{\rho}A\left[\nabla_0\cdot\pmb{\xi}_2+\pmb{\xi}_0\cdot\nabla_0\left(3\epsilon+a\pdn{\epsilon}{a}{}\right)\right]
+i\sigma_0\pmb{D}^{(0)}[\pmb{\xi}_2]+i\sigma_2\pmb{D}^{(0)}[\pmb{\xi}_0]+i\sigma_0\pmb{D}^{(2)}[\pmb{\xi}_0]
\nonumber \\
+\frac{\left(\pmb{\xi}_0\cdot\nabla_0\ln{\rho}+\nabla_0\cdot\pmb{\xi}_0\right)}{4\pi\rho}\left(\nabla_0\times\pmb{B}\right)\times\pmb{B} 
+\frac{1}{4\pi\rho}\left[(\nabla\times\pmb{B}^\prime)\times\pmb{B}+(\nabla\times\pmb{B})\times\pmb{B}^\prime\right] \,, 
\end{eqnarray}
for order $\omega_A^2$. Note that Equation (C6) for order $\omega_A^0$ describes the oscillation of the unmagnetized slowly rotating star. 
The functions $\chi_0$ and $\chi_2$, and $\pmb{B}^\prime$ are, in terms of $\pmb{\xi}_0$ and $\pmb{\xi}_2$, given by 
\begin{eqnarray}
\chi_0&=&-{p\Gamma_1\over\rho}\left( \nabla_0\cdot\pmb{\xi}_0+\pmb{\xi}_0\cdot\nabla_0\ln\rho-\pmb{\xi}_0\cdot\pmb{e}_aA\right)\,,\nonumber \\
\chi_2&=&-{p\Gamma_1\over\rho}\left\{ \nabla_0\cdot\pmb{\xi}_2+\pmb{\xi}_0\cdot\nabla_0\left(3\epsilon+a\pdn{\epsilon}{a}{}\right)
+\pmb{\xi}_2\cdot\nabla_0\ln\rho-\pmb{\xi}_2\cdot\pmb{e}_aA
\right\}\,, \nonumber 
\end{eqnarray}
\begin{eqnarray}
\left(B'\right)^i=\frac{1}{a^2\sin\theta}\epsilon^{ijk}\frac{\partial}{\partial x^j}\left(a^2\sin\theta\,\epsilon_{lmk}\xi_0^lB^m\right)\,. 
\end{eqnarray}
Multiplying Equation (C7) by the complex conjugate of the displacement vector $\pmb{\xi}_0^*$ and
integrating over mass, we obtain the integral relation that includes no term related to $\pmb{\xi}_2$, given by 
\begin{eqnarray}
-2\sigma_0^2\int_0^M\epsilon\left|\pmb{\xi}_0\right|^2\rd
M_a-\sigma_0^2\int_0^M(a\xi^a_0\nabla_0\epsilon)\cdot\pmb{\xi}_0^*\rd
M_a-\sigma_0^2\int_0^M[(\pmb{\xi}_0\cdot\nabla_0\epsilon)\pmb{e}_a]\cdot\pmb{\xi}_0^*\rd
M_a-2\sigma_0\sigma_2\int_0^M\left|\pmb{\xi}_0\right|^2\rd M_a
\nonumber \\
=\int_0^M \chi_0^*\, \pmb{\xi}_0\cdot\nabla_0\left(3\epsilon+a\pdn{\epsilon}{a}{}\right) \rd
M_a+i\sigma_0\int_0^M\pmb{D}^{(2)}[\pmb{\xi}_0]\cdot\pmb{\xi}_0^*\rd
M_a+i\sigma_2\int_0^M\pmb{D}^{(0)}[\pmb{\xi}_0]\cdot\pmb{\xi}_0^*\rd M_a \nonumber \\
+\frac{1}{4\pi}\int_0^M\frac{1}{\rho}\left(-{\rho\over p\Gamma_1}\chi_0+\pmb{\xi}_0\cdot\pmb{e}_aA\right)
\left[\left(\nabla_0\times\pmb{B}\right)\times\pmb{B} \right]\cdot\pmb{\xi}_0^*\rd M_a
+\frac{1}{4\pi}\int_0^M\frac{1}{\rho}\left[(\nabla\times\pmb{B}^\prime)\times\pmb{B}+(\nabla\times\pmb{B})\times\pmb{B}^\prime\right]\cdot\pmb{\xi}_0^*\rd M_a,
\end{eqnarray}
where $\rd M_a=\rho(a) a^2 \sin\theta \rd a \rd\theta\rd\phi$.
Substituting the $\omega_A^0$--order eigenfunctions expanded like Equations (29)-(35) into Equation (C9) and taking
$\sigma_2=E_2^\prime\bar\omega_A^2$, we may obtain the integral expression for the coefficient $E_2^\prime$, given by 
\begin{eqnarray}
E'_2=-\left[\frac{\Omega_{\rm K}^2}{4 \sigma_0}\int_0^Rf_1(a)\hat{\rho}\rho a^4\rd a+\sigma_0\int_0^Rf_2(a)\rho a^4\rd  a
+\frac{\Omega_{\rm K}^2}{2\sigma_0}\int_0^R\frac{1}{c_1}\,f_3(a)\rho a^4\rd a+\Omega \int_0^Rf_4(a)\rho a^4\rd a\right]\bigg/W_I,
\end{eqnarray}
where
\begin{eqnarray}
W_I&=&\int_0^R\left[|\pmb{S}|^2+\pmb{H}^\dagger\pmbmt{\Lambda}_0\,\pmb{H}+\pmb{T}^\dagger\pmbmt{\Lambda}_1\,\pmb{T}\right]\rho\, a^4\rd a  \nonumber \\ 
&&-\frac{\Omega}{\sigma_0}\int_0^R \left[m\left(\pmb{S}^\dagger\pmb{H}+\pmb{H}^\dagger\pmb{S}+|\pmb{H}|^2+|\pmb{T}|^2\right)
+\pmb{S}^\dagger\pmbmt{C}_0\,\pmb{T}-\pmb{T}^\dagger\pmbmt{\Lambda}_1\pmbmt{K}\,\pmb{S}
+\pmb{H}^\dagger\pmbmt{\Lambda}_0\pmbmt{M}_1\,\pmb{T}+\pmb{T}^\dagger\pmbmt{\Lambda}_1\pmbmt{M}_0\,\pmb{H}\right]\rho\,a^4\rd a \,,
\end{eqnarray}
\begin{eqnarray}
f_1(a)=\pmb{S}^\dagger\left\{ma\dfn{\pmb{h}^H}{a}{}-m\pmb{h}^S+2m \left(2+\dfn{\ln\rho}{\ln a}{}\right)\pmb{h}^H
-\pmbmt{C}_0\,\left[a\dfn{\pmb{h}^T}{a}{}+2\left(2+\dfn{\ln\rho}{\ln a}{}\right)\pmb{h}^T\right]\right\}  \nonumber \\
+\pmb{H}^\dagger\left[m\left(2+\dfn{\ln\rho}{\ln a}{}\right)\pmb{h}^S+2m\pmb{h}^H-\pmbmt{\Lambda}_0\pmbmt{R}\,\pmb{h}^T\right]
\nonumber \\
+\pmb{T}^\dagger\left[-\left(2+\dfn{\ln\rho}{\ln a}{}\right)\pmbmt{\Lambda}_1\pmbmt{K}\,\pmb{h}^S+2\pmbmt{\Lambda}_1\pmbmt{M}_0\,\pmb{h}^H
+m(\pmbmt{\Lambda}_1-2\pmbmt{I})\pmb{h}^T\right] 
\nonumber \\
-{2a A\over 3} \left(2+\dfn{\ln\rho}{\ln a}{}\right) \pmb{S}^\dagger\left(\pmbmt{I}-\pmbmt{{\cal{A}}}_0\right) \pmb{S}
+{2V_G\over3}\left(2+\dfn{\ln\rho}{\ln a}{}\right) \pmb{S}^\dagger\left(\pmbmt{I}-\pmbmt{{\cal{A}}}_0\right)\pmb{y}_2
\nonumber \\
+2aA \pmb{H}^\dagger\left(2\pmbmt{{\cal{A}}}_0+\pmbmt{{\cal{B}}}_0\right)\pmb{S}
-2V_G \pmb{H}^\dagger\left(2\pmbmt{{\cal{A}}}_0+\pmbmt{{\cal{B}}}_0\right)\pmb{y}_2
-2maA \pmb{T}^\dagger\pmbmt{Q}_1\pmb{S}
+2mV_G \pmb{T}^\dagger\pmbmt{Q}_1\pmb{y}_2
\,,
\end{eqnarray}
\begin{eqnarray}
f_2(a)=\pmb{S}^\dagger\left[\eta(\bar\alpha)\pmbmt{I}+\eta(\bar\beta)\pmbmt{{\cal{A}}}_0\right]\pmb{S}
+\bar\alpha\left(\pmb{H}^\dagger\pmbmt{\Lambda}_0\,\pmb{H}+\pmb{T}^\dagger\pmbmt{\Lambda}_1\,\pmb{T}\right)
+\bar\beta\,\pmb{H}^\dagger\left(\pmbmt{{\cal{A}}}_0\pmbmt{\Lambda}_0+3\pmbmt{{\cal{B}}}_0\right)\pmb{H}
+\bar\beta\,\pmb{T}^\dagger(\pmbmt{{\cal{A}}}_1\pmbmt{\Lambda}_1+3\pmbmt{{\cal{B}}}_1)\,\pmb{T}
\nonumber \\
-{3\over 2}\bar\beta\pmb{S}^\dagger\pmbmt{{\cal{B}}}_0\,\pmb{H} +{3\over 2}\bar\beta\pmb{H}^\dagger\left(2\pmbmt{{\cal{A}}}_0+\pmbmt{{\cal{B}}}_0\right)\pmb{S} 
+3m \bar\beta\left(\pmb{H}^\dagger\pmbmt{Q}_0\,\pmb{T}+\pmb{T}^\dagger\pmbmt{Q}_1\,\pmb{H}\right)
-{3\over 2}m\bar\beta\left( \pmb{S}^\dagger\pmbmt{Q}_0\,\pmb{T}+\pmb{T}^\dagger\pmbmt{Q}_1\,\pmb{S}\right)  ,
\end{eqnarray}
\begin{eqnarray}
f_3(a)=\pmb{y_2}^\dagger\left[a\dfn{\vartheta(\bar\alpha)}{a}{}\pmbmt{I}+a\dfn{\vartheta(\bar\beta)}{a}{}\pmbmt{{\cal{A}}}_0\right]\pmb{S}
-3\vartheta(\bar\beta)\pmb{y_2}^\dagger\pmbmt{{\cal{B}}}_0\,\pmb{H}-3m\vartheta(\bar\beta)\pmb{y_2}^\dagger\pmbmt{Q}_0\,\pmb{T} \,,
\end{eqnarray}
\begin{eqnarray}
f_4(a)&=&-m \pmb{S}^\dagger \left\{\left[\bar\alpha+\eta(\bar\alpha)\right]\pmbmt{I}+[\bar\beta+\eta(\bar\beta)]\pmbmt{{\cal{A}}}_0\right\}\pmb{H}
-m \pmb{H}^\dagger \left\{\left[\bar\alpha+\eta(\bar\alpha)\right]\pmbmt{I}+[\bar\beta+\eta(\bar\beta)]\pmbmt{{\cal{A}}}_0\right\}\pmb{S}
\nonumber \\
&&-2m\bar\alpha\left(|\pmb{H}|^2+|\pmb{T}|^2\right)
-4\bar\beta\,\pmb{T}^\dagger\left(\pmbmt{{\cal{A}}}_1\pmbmt{\Lambda}_1\pmbmt{M}_0+3\pmbmt{Q}_1\pmbmt{{\cal{B}}}_0\right)\pmb{H}
\nonumber \\
&&-\pmb{S}^\dagger \left\{[\bar\alpha+\eta(\bar\alpha)]\pmbmt{C}_0+[\bar\beta+\eta(\bar\beta)]\pmbmt{{\cal{A}}}_0\pmbmt{C}_0\right\}\pmb{T}
-2(\bar\alpha-\bar\beta)\,\pmb{H}^\dagger\pmbmt{\Lambda}_0\pmbmt{M}_1\,\pmb{T}
\nonumber \\
&&-2(\bar\alpha-\bar\beta)\,\pmb{T}^\dagger\pmbmt{\Lambda}_1\pmbmt{M}_0\,\pmb{H}
-2m\bar\beta\,\pmb{H}^\dagger\left(\pmbmt{I}+6\pmbmt{{\cal{A}}}_0\right)\pmb{H}
-4\bar\beta\,\pmb{H}^\dagger \left(\pmbmt{{\cal{A}}}_0\pmbmt{\Lambda}_0\pmbmt{M}_1+3\pmbmt{Q}_0\pmbmt{{\cal{B}}}_1\right)\pmb{T}
\nonumber \\
&&-2m\bar\beta\,\pmb{T}^\dagger\left(\pmbmt{I}+6\pmbmt{{\cal{A}}}_1\right)\pmb{T}
+\left[\bar\alpha+\eta(\bar\alpha)\right]\pmb{T}^\dagger\pmbmt{\Lambda}_1\pmbmt{K}\,\pmb{S}
+\left[\beta+\eta(\beta)\right]\pmb{T}^\dagger\left(\pmbmt{{\cal{A}}}_1\pmbmt{\Lambda}_1\pmbmt{K}+3\pmbmt{Q}_1\pmbmt{Q}_0\pmbmt{Q}_1-3\pmbmt{Q}_1\right)\pmb{S} \,. 
\end{eqnarray}
Here, $\bar{\alpha}\equiv\alpha/\bar{\omega}_A^2$, $\bar{\beta}\equiv\beta/\bar{\omega}_A^2$, $\Omega_{\rm  K}\equiv(GM/R^3)^{1/2}$, and 
$\pmb{X}^\dagger$ means the Hermitian conjugate of the complex matrix $\pmb{X}$. 
The magnetic perturbations $\pmb{h}^S$, $\pmb{h}^H$, and $\pmb{h}^T$ are given by 
\begin{eqnarray}
\pmb{h}^S=m\pmb{S} \,,
\end{eqnarray}
\begin{eqnarray}
\pmb{h}^H=maA\,\pmbmt{\Lambda}_0^{-1}\pmb{S}-mV_G\pmbmt{\Lambda}_0^{-1}\pmb{y_2}+m\pmb{H} \,,
\end{eqnarray}
\begin{eqnarray}
\pmb{h}^T=aA\,\pmbmt{K}\pmb{S}-V_G\pmbmt{K}\,\pmb{y_2}-m\pmb{T} \,.
\end{eqnarray}
In the expression for $E'_2$ given above, all the eigenfunctions, $\pmb{S}$, $\pmb{H}$, $\pmb{T}$, and $\pmb{y}_2$ are for unmagnetized stars 
even though the subscript ``0'' is not attached. 

When inertial modes are considered for the $\omega_A^0$--order eigensolution, for which 
$\displaystyle \lim_{\Omega\rightarrow 0} {\sigma_0\over\Omega}=\kappa_0$ with $\kappa_0$ being a constant, from Equation (C10), we see that  
\begin{eqnarray}
{\sigma_2\over\sigma_0}\rightarrow\frac{\eta_2'}{\bar{\Omega}^2}\bar\omega_A^2 \quad {\rm as} \ \bar{\Omega}\rightarrow 0 \,,
\label{s2s0}
\end{eqnarray}
where $\eta_2'$ is a constant depending on the mode considered, which is given by 
\begin{eqnarray}
\eta_2'=-\lim_{\bar\Omega\to 0}{1\over 4\kappa_0^2\,W_{I}}\left[\int_0^Rf_1(a)\hat{\rho}\rho a^4\rd a+2\int_0^R\frac{1}{c_1}f_3(a)\rho a^4\rd a\right] \,. 
\end{eqnarray}
Equation (\ref{s2s0}) implies that for the inertial mode, our expression 
for $\sigma_2$ becomes inappropriate in the case of $\bar{\Omega}^2 \lse \bar\omega_A^2$. For the inertial mode, therefore, 
the condition $\bar\omega_A^2 \ll \bar{\Omega}^2 \ll 1$ is required
for the expression  for $\sigma_2$ to be applicable. Similar expressions to Eq.~(\ref{s2s0}) but for stars with general magnetic field distribution have been obtained by Morsink \& Rezania (2002).


\label{lastpage}


\begin{thebibliography}{99}
\bibitem[\protect\citeauthoryear{Andersson}{2009}] {} Andersson N.,
  Glampedakis K., Samuelsson L., 2009, MNRAS, 396, 894

\bibitem[\protect\citeauthoryear{Arras et al}{2004}]{b1} Arras P.,
Cumming A., Thompson C., 2004, ApJ, 608, L49

\bibitem[\protect\citeauthoryear{AsaiLee}{2014}] {} Asai H., Lee U., 2014, ApJ, 790, 66

\bibitem[\protect\citeauthoryear{Braithwaite \& Spruit}{2004}]{b22}
  Braithwaite J., Spruit H. C., 2004, Nature, 431, 819
  
\bibitem[\protect\citeauthoryear{Cerd$\acute{{\rm{a}}}$-Dur$\acute{{\rm{a}}}$n
  et al}{2009}]{b22} Cerd$\acute{{\rm{a}}}$-Dur$\acute{{\rm{a}}}$n P.,
Stergioulas N., Font J. A., 2009, MNRAS, 397, 1607

\bibitem[\protect\citeauthoryear{Ciolfi et al}{2009}]{b1} Ciolfi R.,
Ferrari V., Gualtieri L., Pons J. A., 2009, MNRAS, 397, 913

\bibitem[\protect\citeauthoryear{Ciolfi \& Rezzolla}{2012}]{b1} Ciolfi R.,
Rezzolla L., 2012, ApJ, 760, 1

\bibitem[\protect\citeauthoryear{Colaiuda et al.}{2008}]{b1} Colaiuda
  A., Ferrari V., Gualtieri L., Pons J. A., 2008, MNRAS, 385, 2080
  
\bibitem[\protect\citeauthoryear{Colaiuda \& Kokkotas}{2011}]{b2} Colaiuda A., Kokkotas K. D., 2011, MNRAS, 414, 3014

\bibitem[\protect\citeauthoryear{Colaiuda \& Kokkotas}{2012}]{b22}
  Colaiuda A., Kokkotas K. D., 2012, MNRAS, 423, 811
  
  \bibitem[]{} Duncan R.C., 1998, ApJL, 498, L45
  
\bibitem[\protect\citeauthoryear{Frieben \& Rezzolla}{2012}]{b1} Frieben, J., 
 Rezzolla, L., 2012, MNRAS, 427, 3406 
 
\bibitem[\protect\citeauthoryear{Gabler et al.}{2011}]{b5} Gabler M.,
  Cerd$\acute{{\rm{a}}}$-Dur$\acute{{\rm{a}}}$n P., Font J. A.,
  M$\ddot{{\rm{u}}}$ller E., Stergioulas N., 2011, MNRAS, 410, L37
  
\bibitem[\protect\citeauthoryear{Gabler et al.}{2012}]{b5} Gabler M.,
  Cerd$\acute{{\rm{a}}}$-Dur$\acute{{\rm{a}}}$n P., Stergioulas N., Font J. A.,
  M$\ddot{{\rm{u}}}$ller E., 2012, MNRAS, 421, 2054
  
\bibitem[\protect\citeauthoryear{Gabler et al.}{2013a}]{b5} Gabler M.,
  Cerd$\acute{{\rm{a}}}$-Dur$\acute{{\rm{a}}}$n P., Font J. A.,
  M$\ddot{{\rm{u}}}$ller E., Stergioulas N., 2013, MNRAS, 430, 1811
 
\bibitem[\protect\citeauthoryear{Gabler et al.}{2013b}]{b5} Gabler M.,
  Cerd$\acute{{\rm{a}}}$-Dur$\acute{{\rm{a}}}$n P., Stergioulas N., Font J. A.,
  M$\ddot{{\rm{u}}}$ller E., 2013, PhRvL, 111
 
 \bibitem[]{} Glampedakis K., Samuelsson L., Andersson N., 2006, MNRAS, 371, L74 
 
  \bibitem[]{} Glampedakis K., Andersson N., Samuelsson L., 2011, MNRAS, 410, 805 
 
\bibitem[\protect\citeauthoryear{Goedbloed \& Poedts}{2004}]{GP04} Goedbloed H., Poedts S., 2004,
Principles of Magnetohydrodynamics with Applications to Laboratory and Astrophysical Plasma,
Cambridge University Press, Cambridge

\bibitem[\protect\citeauthoryear{Goosens}{1979}]{b1} Goosens M., 1979,
  A\&A, 123,147
  
\bibitem[\protect\citeauthoryear{Hambaryan et al.}{2011}]{b3}
  Hambaryan V., Neuh$\ddot{\rm{a}}$user R., Kokkotas K. D., 2011,
  A\&A, 528, A45
  
  \bibitem[]{} Huppenkothen D., et al, 2014, ApJ, 787, 128
  
\bibitem[\protect\citeauthoryear{Israel et al.}{2005}]{b7} Israel G.,
  Belloni T., Stella L., Rephaeli Y., Gruber D. E., Casella P.,
  Dall'Osso S., Rea N., Persic M., Rothschild R. E., 2005, ApJ, 628,
  L53
  
\bibitem[\protect\citeauthoryear{Kiuchi \& Yoshida}{2008}]{b1} Kiuchi K.,
Yoshida S., 2008, PRD, 78, 044045

\bibitem[\protect\citeauthoryear{Kiuchi et al.}{2011}]{b1} 
Kiuchi, K., Yoshida, S., Shibata, M., 2011, A\&A, 532, A30

\bibitem[\protect\citeauthoryrar{Kotate et al}{2006}]{b1} Kotate, K.,
  Sato, K., Takahashi, K. 2006, Reports of Progress in Physics, 69, 971
  
\bibitem[\protect\citeauthoryear{Lander \& Jones}{2011a}]{b1} Lander
  S. K., Jones D. I., 2011, MNRAS, 412, 1394

\bibitem[\protect\citeauthoryear{Lander \& Jones}{2011b}]{b1} Lander
  S. K., Jones D. I., 2011, MNRAS, 412, 1730

\bibitem[]{} Lander S. K., Jones D.I., Passamonti A., 2010, MNRAS,
  405, 318
  
\bibitem[\protect\citeauthoryear{Lasky et al}{2011}]{b1} Lasky P. D.,
Zink B., Kokkotas K. D., Glampedakis K., 2011, ApJ, 735, L20

\bibitem[\protect\citeauthoryear{Lee}{1993}]{b9} Lee U., 1993, ApJ,
  405, 359
\bibitem[\protect\citeauthoryear{Lee}{2005}]{b9} Lee U., 2005, MNRAS, 357, 97

\bibitem[\protect\citeauthoryear{Lee}{2007}]{b9} Lee U., 2007, MNRAS, 374, 1015

\bibitem[]{} Lee U., 2008, MNRAS, 385, 2069

\bibitem[]{} Levin Y., 2006, MNRAS, 368, L35

\bibitem[]{} Levin Y., 2007, MNRAS, 377, 159

\bibitem[]{} Lockitch, K. H., Friedman, J. L., 1999, ApJ, 521, 764

\bibitem[\protect\citeauthoryear{Mathis \& Brye}{2011}]{b9} Mathis S.,
  Brye N., 2011, A\&A, 526, A65
  
\bibitem[\protect\citeauthoryear{Mathis \& Brye}{2012}]{b9} Mathis S.,
  Brye N., 2012, A\&A, 540, A37
  
  \bibitem[\protect\citeauthoryear{mereghetti}{2008}]{mer08} Mereghetti S., 2008, Astron. Astrophys. Rev., 15, 225
  
\bibitem[\protect\citeauthoryear{Miketinac}{1973}]{b9} Miketinac, M. J., 1973, Ap\&SS, 22, 413

\bibitem[]{} Morsink S. M., Rezania V., 2002, ApJ, 574, 908
  
  \bibitem[]{} Passamonti A., Lander S.K., 2013, MNRAS, 429, 767

  \bibitem[]{} Passamonti A., Lander S.K., 2014, MNRAS, 438, 156
  
  \bibitem[]{} Piro A.L., 2005, ApJ, 634, L153
  
\bibitem[\protect\citeauthoryear{Saio}{1981}]{b15} Saio H., 1981, ApJ,
  244, 299
  
\bibitem[\protect\citeauthoryear{Smeyers \& Denis}{1971}]{b15} Smeyers
  P., Denis J., 1971, A\&A, 14, 311

\bibitem[]{} Sotani H., Kokkotas K. D., Stergioulas N., 2007, MNRAS,
  375, 261
  
\bibitem[\protect\citeauthoryear{Sotani, Kokkotas \&
    Stergioulas}{2008}]{b18} Sotani H., Kokkotas K. D., Stergioulas
  N., 2008, MNRAS, 385, L5
  
\bibitem[\protect\citeauthoryear{Sotani, Colaiuda \&
    Kokkotas}{2008}]{b18} Sotani H., Colaiuda A., Kokkotas K. D.,
  2008, MNRAS, 385, 2161
  
\bibitem[\protect\citeauthoryear{Strohmayer \& Watts}{2005}]{b19} Strohmayer T. E., Watts A. L., 2005, ApJ, 632, L111

\bibitem[\protect\citeauthoryear{Strohmayer \& Watts}{2006}]{b20}
  Strohmayer T. E., Watts A. L., 2006, ApJ, 653, 593
  
  \bibitem[]{} Thompson C., Duncan R.C., 1993, ApJ, 408, 194
  
  
  \bibitem[]{} Thompson C., Duncan R.C., 1996, ApJ, 473, 322

  \bibitem[]{} Thompson C., Lyuitikov M., Kulkarni S.R., 2002, ApJ, 574, 332 
  
  
\bibitem[\protect\citeauthoryear{Unno etal}{1989}]{Unno1} Unno, W., Osaki, Y., Ando, H., Saio, H, Shibahashi, H., 1989, Nonradial Oscillations of Stars, 2nd Ed, University of Tokyo Press, Tokyo

\bibitem[]{} van Hoven M.B., Levin Y., 2011, MNRAS, 410, 1036

\bibitem[]{} van Hoven M.B., Levin Y., 2012, MNRAS, 420, 3035

\bibitem[]{} Watts A.L., Strohmayer T.E., 2006, ApJ, 637, L117

\bibitem[]{} Watts A.L., 2011, arXiv:1111.0514v1

\bibitem[\protect\citeauthoryear{woodsThompson}{2006}] {} Woods P.M., Thompson C., 2006, in Compact stellar X-ray sources, 
ed. W.H.G. Lewin \& M. van der Klis (Cambrisge: Cambridge Univ. Press), p547

\bibitem[\protect\citeauthoryear{Yoshida \& Eriguchi}{2006}]{b1} 
Yoshida, S., Eriguchi, Y., 2006, ApJS, 164, 156

\bibitem[\protect\citeauthoryear{Yoshida \& Lee}{2000a}]{b1} Yoshida
  S., Lee U., 2000a, ApJ, 529, 997
  
\bibitem[\protect\citeauthoryear{Yoshida \& Lee}{2000b}]{b1} Yoshida
  S., Lee U., 2000b, ApJS, 120, 353
  
\bibitem[\protect\citeauthoryear{Yoshida et al}{2006}]{b1} Yoshida S.,
Yoshida S., Eriguchi Y., 2006, ApJ, 651, 462
\end{thebibliography}
\end{document}